\documentclass[journal,twoside,web]{IEEEtran}
\usepackage{cite}
\usepackage{amsmath,amssymb,amsfonts}
\usepackage{algorithmic}
\usepackage{graphicx}
\usepackage{resizegather}
\usepackage{booktabs}
\usepackage[table,xcdraw]{xcolor}
\usepackage{resizegather}
\usepackage{booktabs}
\usepackage{hyperref}
\usepackage{caption}
\def\BibTeX{{\rm B\kern-.05em{\sc i\kern-.025em b}\kern-.08em
		T\kern-.1667em\lower.7ex\hbox{E}\kern-.125emX}}
\begin{document}
	\title{Homodyned K-Distribution Parameter Estimation in Quantitative Ultrasound: Autoencoder and Bayesian Neural Network Approaches}
	\author{Ali K. Z. Tehrani, Guy Cloutier, An Tang,\\  Ivan M. Rosado-Mendez$^*$ and Hassan Rivaz$^*$ 
		\thanks{A. K. Z. Tehrani and H. Rivaz are with the Department
			of Electrical and Computer Engineering, Concordia University, QC,  Canada.
			An Tang is with the Department of Radiology, Radio-oncology and Nuclear Medicine, University of Montreal, QC, Canada.	
			Guy Cloutier is with the Department of Radiology, Radio-oncology and Nuclear Medicine, and Institute of Biomedical Engineering, University of Montreal, QC, Canada.	
			Ivan M. Rosado-Mendez is with the Department of Medical Physics and Radiology, University of Wisconsin, United States.
			e-mail: A\_Kafaei@encs.concordia.ca, rosadomendez@wisc.edu, an.tang@umontreal.ca, guy.cloutier@umontreal.ca, and  
			hrivaz@ece.concordia.ca. $^*$ represents joint senior authorship with equal contribution.  }%
		\thanks{}}
	
	\maketitle
	
	\begin{abstract}
		Quantitative ultrasound (QUS) analyzes the ultrasound backscattered data to find the properties of scatterers that correlate with the tissue microstructure. Statistics of the envelope of the backscattered radiofrequency (RF) data can be utilized to estimate several QUS parameters. Different distributions have been proposed to model envelope data. The homodyned K-distribution (HK-distribution) is one of the most comprehensive distributions that can model ultrasound backscattered envelope data under diverse scattering conditions (varying scatterer number density and coherent scattering). The scatterer clustering parameter ($\alpha$) and the ratio of the coherent to diffuse scattering power ($k$) are the parameters of this distribution that have been used extensively for tissue characterization in diagnostic ultrasound. The estimation of these two parameters (which we refer to as HK parameters) is done using optimization algorithms in which statistical features such as the envelope point-wise signal-to-noise ratio (SNR), skewness, kurtosis, and the log-based moments have been utilized as input to such algorithms. The optimization methods minimize the difference between features and their theoretical value from the HK model. We propose that the true value of these statistical features is a hyperplane that covers a small portion of the feature space. \textcolor{black}{In this paper, we follow two approaches to reduce the effect of sample features' error. We propose a model projection neural network based on denoising autoencoders to project the noisy features into this space based on this assumption. We also investigate if the noise distribution can be learned by the deep estimators.} We compare the proposed methods with conventional methods using simulations, an experimental phantom, and data from an \textit{in vivo} animal model of hepatic steatosis. \textcolor{black}{The network weight and a demo code are available online at \href{http://code.sonography.ai}{http://code.sonography.ai}}     
	\end{abstract}
	
	
	\section{Introduction}
	\label{sec:introduction}
	Scatterers are microstructures that scatter ultrasound waves and are typically smaller than the wavelength of the ultrasound wave. Quantitative ultrasound (QUS) tries to provide insight into the scatterers' characterisitcs from the analysis of the detected backscattered signals \cite{oelze2016review,wagner1983statistics}. QUS methods can be broadly classified into three categories: spectral-based, motion-based, and time-domain-based methods. Spectral-based methods utilize backscattered Radio frequency (RF) data compression wave to estimate parameters like the backscatter coefficient and the attenuation coefficient, while also removing system-dependent effects through the use of a reference phantom \cite{jafarpisheh2020analytic,Vajihi2018,yao1990backscatter,rouyer2016vivo,soylu2023calibrating}. Motion-based QUS methods study the behavior of scatterers in response to an induced motion, measuring parameters such as the elastic modulus \cite{mohammadi2021ultrasound,tehrani2020displacement} and the shear wave attenuation reflecting tissue viscosity \cite{yazdani2022revisited}. Finally, time-domain-based methods estimate parameters related to the number and coherency of scatterers by fitting a distribution to the envelope of backscattered RF data \cite{rosado2016analysis,destrempes2013review,dutt1995speckle}. Spectral-based analysis of the backscatter coefficient and time-domain analysis of the echo envelope share common and complementary information \cite{destrempes2015unifying}.
	
	The Nakagami and Homodyned K-distributions (HK-distribution) are the probability density functions that have been used more frequently to model the envelope data, and the parameters of these distributions are found to be useful in tissue characterization including hepatic steatosis grading \cite{destrempes2022quantitative,fang2020ultrasound,zhou2019hepatic,zhou2020value,nguyen2019reference}, breast cancer diagnosis and characterization \cite{muhtadi2022breast,chowdhury2022ultrasound,destrempes2020added,trop2015added,byra2016classification} and classification of metastatic lymph nodes \cite{hoerig2023classification}. The HK-distribution is a more comprehensive distribution compared to the Nakagami distribution due to the fact that it can model and differentiate scattering with a high number of scatterers per resolution cell where the Nakagami fails \cite{zhou2020value}. However, the HK-distribution requires a larger number of samples of the amplitude of the detected echo signals to achieve similar levels of accuracy and precision in parameter estimation  compared to the Nakagami distribution.
	
	The parameters of the HK distribution, referred here to as the HK parameters, are also physically meaningful. The scatterer clustering parameter ($\alpha$) and the ratio of the coherent-to-diffuse scattering power ($k$) are two parameters of this distribution that are related to the scatterer number density, and the microstructural organization of scatterers, respectively.  
	Envelope statistics such as point-wise signal-to-noise ratio (SNR), skewness, kurtosis, and the log-based moments are usually employed to estimate HK parameters. Hruska \textit{et al. } employed SNR, skewness, and kurtosis to estimate the HK parameters by minimizing the difference between their sample estimates and the theoretical values \cite{Hruska2009}. Destrempes \textit{et al.} proposed to employ two log-based moments $X$ and $U$ and used bi-section interpolation to estimate the parameters \cite{destrempes2013estimation}. They reported reduction of bias and variance by using these two moments (we refer to this method as the XU method). Liu \textit{et al.} proposed to utilize several statistics (in total 16) and compared them with the theoretical values using table search \cite{liu2023study}. They also performed an extensive analysis on feature selection for each HK parameter estimation.

	\textcolor{black}{One of the constraints inherent in conventional methods is due to their high computational complexity, which limits their translation into real-time \textit{in vivo} imaging applications. Recently, deep learning (DL) methods are being used more frequently for HK parameter estimation. These method can be implemented efficiently on GPU which can obtain real-time performance. Zhou \textit{et al.} used a multilayer perceptron (MLP) that takes envelope statistics as input, and outputs the HK parameters \cite{zhou2021parameter}. We also demonstrated that deep methods can be utilized for not only fast estimation of the parameters, but also to quantify the uncertainty of the estimated parameter \cite{tehrani2022homodyned}.}
	
	\textcolor{black}{The HK-distribution parametric images can be formed by spatially dividing the envelope data into small grids. A patch around the center of each grid is selected to estimate HK parameters. A larger patch size provides a more reliable estimate. However, a larger patch size increases the heterogeneity within the patch, potentially resulting in the optimization methods' failure. Smaller patch size results in deviation of computed parameters from the theoretical values, leading to errors in estimating HK parameters.}
	
	In this paper, we focus on reducing the estimation error due to low sample size. To achieve this, we consider the feature space as a high-dimensional space where each patch is represented by a point. We show that the envelope statistics features lie in a hyperplane that covers only a small volume of the feature space, and hypothesize that noisy estimates of the statistical features of the envelope deviate from this hyperplane because of statistical errors. We then propose to project the points into this hyperplane using an autoencoder to reconstruct clean features from noisy sample estimates. The reconstructed features can be utilized to estimate the HK-distribution parameters with either non-DL (like the XU estimator) or DL methods (MLP, BNN). 
	
	In addition to the autoencoder approach, we expand the investigation of our recently proposed Bayesian neural network (BNN) \cite{tehrani2022homodyned} because of its ability to provide estimates of the uncertainty of parameter estimation. This uncertainty can be of clinical value to convey clinicians the confidence on the HK parameters. We investigate different training strategies of the proposed BNN for different sample sizes.

	The two approaches are comprehensively compared with conventional estimators and validated using simulations, experimental phantoms, and \textit{in vivo} data.

	\section{Materials and Method}
	\subsection{Homodyned K-distribution Parameters Estimation Problem Formulation}

	\subsubsection{Homodyned K-distribution}
	The Homodyned K-distribution can be defined as \cite{destrempes2013estimation}:
	\begin{equation}
		P_{HK}(A|\epsilon ,\sigma^2,\alpha) = A\int_{0}^{\infty}uJ_{0}(u\epsilon)J_{0}(uA)(1+\frac{u^2\sigma^2}{2})^{-\alpha}du
	\end{equation}
	where $A$ is the envelope of the backscattered echo signal, $\alpha$ is the scatterer clustering parameter that is related to the scatterer number density, and $J_{0}(.)$ denotes the zero-order Bessel function. The coherent signal power is represented by $\epsilon^2$, and the diffuse signal power is $2\sigma^2\alpha$ \cite{destrempes2013estimation}. The scatterer clustering parameter ($\alpha$), and coherent to diffuse scattering ratio $k = \frac{\epsilon}{\sigma \sqrt{\alpha}}$ are the HK-distribution parameters that are commonly used in tissue characterization. The envelope statistics (SNR, skewness, kurtosis of the fractional amplitude, and log-based moments) can be obtained by:
	
	\begin{equation}
		\label{eq:feature}		
		\begin{gathered}		
			SNR (R^v)=\frac{E[A^v]}{\sqrt{E[A^{2v}]-(E[A^v])^2}},\\ \\
			Skewness (S^v)= \frac{E[(A^v-E[A^v])^3]}{(E[A^{2v}]-(E[A^v])^2)^{1.5}},\\ \\
			Kurtosis (K^v) = \frac{E[A^{4v}]-4E[A^v] \times E[A^{3v}] + 6E[A^{2v}] \times E[A^{v}]^2-3E[A^{v}]^4}{(E[A^{2v}] -E[A^{v}]^2)^2},\\ \\
			U = E[log(I)] - log(E[I]), \\ \\
			X = E[I\times log(I)]/E[I]-E[log(I)],
		\end{gathered}
	\end{equation}
	where $I=A^2$ and $v$ takes the two values ${0.72,0.88}$ as recommended by \cite{Hruska2009}.

	\subsubsection{HK parameters inverse problem}
	Let the envelope statistic features from a single patch be the vector, $F\in \mathbb{R}^{M\times 1} = [ R^{0.72}, R^{0.88}, S^{0.72}, S^{0.88}, K^{0.72}, K^{0.88}, X, U ]^T$, where $M$ is the number of envelope statistic features fixed at 8 similar to \cite{zhou2021parameter}, and the vector of HK parameters be denoted as $\Theta\in \mathbb{R}^{2\times 1}=[\log_{10}(\alpha),k]^T$. The forward and inverse problems can be illustrated as:
	\begin{equation}
		\begin{gathered}
			\Theta \in \mathbb{R}^{2\times 1}\xrightarrow[]{forward}F\in \mathbb{R}^{M\times 1},\\
			\widetilde{\Theta} \in \mathbb{R}^{2\times 1}\xleftarrow[]{inverse}\widetilde{F}\in \mathbb{R}^{M\times 1},\\
		\end{gathered}
	\end{equation}
	where $\widetilde{F}$ denotes the sample estimate of the envelope statistics, and $\widetilde{\Theta}$ represents the estimated $\Theta$. 
	The forward problem can be viewed as obtaining the envelope statistics from the known parameters of the distribution. Hruska \textit{et al.} showed that moments of the HK-distribution can be analytically obtained from a known HK-distribution by \cite{Hruska2009}:
	\begin{equation}
		\label{eq:anal}
		\begin{gathered}
			E[A^v]=\int_{0}^{\infty}(\frac{2\sigma^2}{\alpha})^{v/2}\frac{\Gamma(1+v/2)x^{v/2+\alpha-1}}{\Gamma(\alpha)e^{x}}\, \\{1}^{}\textrm{F}_1(-v/2;1;\frac{-\alpha \varepsilon^2}{2\sigma^2 x})dx,
		\end{gathered}
	\end{equation}
	where ${1}^{}\textrm{F}_1(a,b,x)$ is a confluent hypergeometric function of the first kind. The feature vector then can be obtained by inserting $E[A^v]$ obtained from Eq \ref{eq:anal} into Eq \ref{eq:feature}. \textcolor{black}{The Eq \ref{eq:anal} is numerically solved by considering discreet values for $x$, and summing the equation inside the integral. For integer values of $x$, the equation is not defined and obtained by interpolation.}

	Estimation of HK parameters from envelope statistics can be viewed as an inverse problem which maps $\widetilde{F} $ from the feature space with $M$ dimensions to the 2-dimensional HK parameter space. The sample mean is used to approximate $E[A^v]$, and the vector with the calculated sample envelope statistic features ($\widetilde{F}$) is employed to estimate HK parameters ($\widetilde{\Theta}$). The low dimensional space of HK parameters enforces the feasible feature values to lie in a low dimensional manifold. Inspired by this, we designed a model projection neural network based on a denoising autoencoder to project the noisy features (\textcolor{black}{$\tilde{F}$}) into the feasible hyperplane.

	\subsection{Model Projection Autoencoder (MPAE)}
\begin{figure}[t]
		
		\centering
		\includegraphics[width=0.40\textwidth]{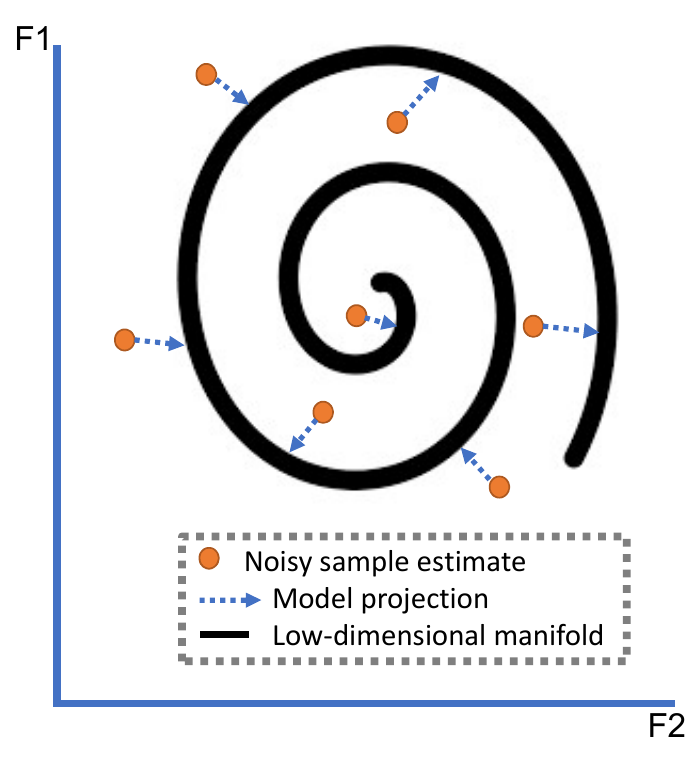}
		\caption{ Graphical representation of the denoising autoencoder. The theoretical values of envelope statistics lie in a low-dimensional manifold. Sample estimates are corrupted by noise and lie further away from the manifold. The MPAE projects the noisy sample estimates into the low-dimensional manifold.}
		\label{fig:manifold}	
	\end{figure}

	Autoencoders are neural networks that receive the input and map them to a lower-dimensional representation by transforming into more informative lower dimensional features. In the encoder part, the input dimension is reduced using several hidden layers, and in the decoder part, the input is reconstructed using the lower-dimensional representation \cite{vincent2008extracting}. These networks have been found useful in many applications, such as denoising and dimensionality reduction, where it was shown that they act as a non-linear principal component analysis (PCA) \cite{ladjal2019pca}. Denoising autoencoders employ the corrupted data and try to reconstruct the clean data. Vincent \textit{et al.} showed that if data lie in a low-dimensional manifold, the corrupted data will be further away from this manifold, and denoising autoencoders can project the corrupted data into the low-dimensional manifold \cite{vincent2008extracting}. This idea is illustrated in Fig. \ref{fig:manifold}.
	
\begin{figure*}[t]	
		\centering
		\includegraphics[width=0.95\textwidth]{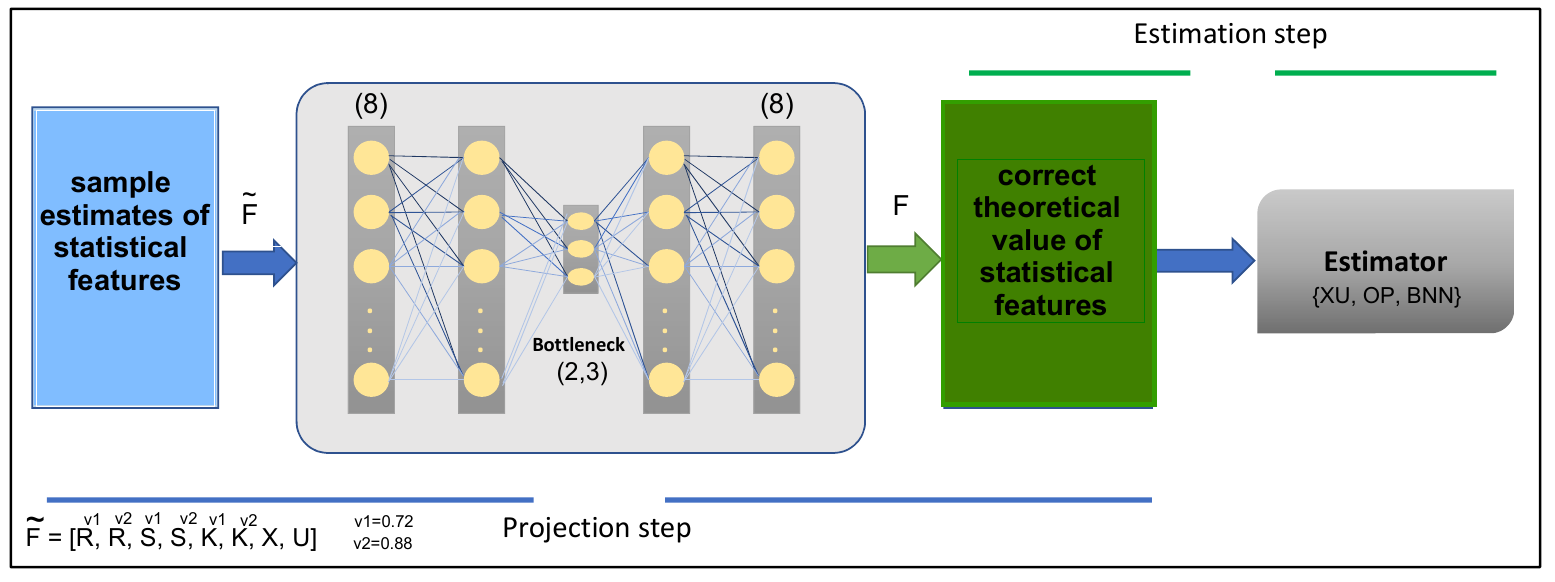}
		\caption{ The proposed two-step framework for estimation of HK parameters. The sample estimate of envelope statistical features ($\tilde{F}$) is projected into a low-dimensional space (bottleneck) and the clear features are used in the estimation of HK parameters (estimation step).}
		\label{fig:network}
	\end{figure*}

	
	Our proposed methods is illustrated in Fig. \ref{fig:network} and has two steps: a projection step and an estimation step. In the projection step, we considered the sample estimates of the statistical features as the corrupted input data and the theoretical value from the forward problem as the clean output features. A model projection autoencoder (MPAE) was utilized to obtain clean features (Projection step). In the estimation step, an estimator (either non-DL or DL-based) was employed to estimate the HK parameters from the denoised features. 
	\subsubsection{Network architecture and training schedule}
	
	The network comprised seven layers; the encoder part had four layers with $64,32,32,b$ nodes, and the decoder had three layers with $32,32,8$ nodes. The parameter $b$ was the number of nodes in the bottleneck, which determined the size of a low-dimensional manifold. We investigated different sizes of bottlenecks in the Results section. The activation functions of all layers were leaky ReLu except the last layer, which did not have any activation function. We tested placing dropouts in different layers and obtained the best performance on placing a dropout on the second encoder layer with the probability of 0.2. \textcolor{black}{We should clarify that the autoencoder inputs are usually high-dimensional data such as image and texture features, hence the networks tend to decrease the feature dimension by reducing the number of nodes. However, in our case, the input has only 8 variables. We expanded it to a higher dimension to increase the learning ability of the network.}

	The features were normalized to have a zero mean and a standard deviation of 1. We employed the combination of a smooth L1-norm (L1) and the mean square error (MSE) as the loss function, which can be given as:
	\textcolor{black}{
		\begin{equation}
			loss = ||F-\tilde{F}||_2 + ||F-\tilde{F}||_{1s}
	\end{equation}}
	\textcolor{black}{	where $||.||_2$ denotes the L2-norm (MSE) and $||.||_{1s}$ represents smooth the L1-norm which is added to penalize small errors as well and being more robust to outliers.}

	For each simulated sample size (the samples drawn from the HK-distribution) of $N_s=4096$, $N_s = 1024$, and $N_s=256$, a separate MPAE was trained since the noise distribution is different for each sample size.       
	\subsection{HK Parameters Estimators}
	We employed BNN as the estimator after MPAE feature reconstruction. The method was also compared with XU estimator and an optimization method.
	
	\subsubsection{XU estimator}
	Destrempes \textit{et al.} proposed to employ two log-based moments, $X$ and $U$ as the statistical features \cite{destrempes2013estimation}. They used a bi-section interpolation to find the intersection between the theoretical values and the sample estimates of $X$ and $U$. They reported improved estimation using these two moments compared to the method proposed by Hruska and Oelze \cite{Hruska2009} that relies on $R$, $S$, and $K$.
	\subsubsection{Optimization method}
	\textcolor{black}{The optimization cost function can be formulated as: $J(\alpha ,k) = arg_{\Theta}\; min\left \{ ||\tilde{F}-F_{(\alpha ,k)}||_2 ) \right \}$, where $\tilde{F}$ is the sample estimate of the statistical features and $F$ denotes the theoretical value. The optimization method can be a simple table search or any other optimization method. We employed particle swarm optimization to find $\alpha$ and $k$, whereas Liu \textit{et al.} utilized table search \cite{liu2023study}.} 
	
	\subsubsection{BNN estimator}
	In \cite{tehrani2022homodyned}, we developed a BNN to estimate the HK parameters. The BNN estimator outperformed the artificial neural network (ANN) counterpart \cite{zhou2021parameter} and was able to quantify the uncertainty in parameter estimation. In BNN, the network weights are sampled from a distribution learned in the training phase. Each time the network runs, the weights are different; therefore, the network was run for each feature vector multiple times (we executed the network 50 times) during the inference stage, and the average value and standard deviations were considered as the prediction and uncertainty, respectively. In this paper, we shed more light one this method by comparing different training strategies to see which method is best fitted to be employed for low-sample sizes.

	\textcolor{black}{Different approaches can be followed in the training of BNN. The theoretical values of the statistical features for different values of the HK parameters can be employed for training the BNN. This method (we named it as BNN-Th) allows us to use one network for all sample sizes, similar to the generalized neural network in \cite{wu2023parallelized}. However, the network is not informed about the noise distribution of the sample input. The proposed MPAE can be utilized to reconstruct the features for this estimator, an approach we named BNN-Th+MPAE. Another approach is to train the BNN using the sample estimate of the features (which we named as BNN-Sam). This method is informed about the noise distribution of the features due to low sample size but for each sample size a different BNN should be used. Another approach to train the BNN is to employ the reconstructed features from MPAE as the feature input for the BNN (which we named as BNN-Sam+MPAE). Although MPAE tries to make the features as close as possible to the theoretical features, noise and variations are present in the reconstructed features especially for very low sample sizes; hence by this method, BNN sees the noise and variations presented in MPAE output during the training.}

	\subsection{Datasets and data generation}
	\subsubsection{Simulation data}
	\textcolor{black}{Training data was generated by sampling from the HK-distribution. The equation suggested by \cite{Hruska2009,zhou2021parameter} was employed to generate the simulation data:}
	\textcolor{black}{\begin{equation}
			a_i = \sqrt{\left (  \sqrt{2k} +X\sigma \sqrt{Z/\alpha }\right )^2+\left (  Y\sigma \sqrt{Z/\alpha }\right )^2}
	\end{equation}}
	
	\textcolor{black}{where $\alpha$ and $k$ are the scatterer clustering parameter and coherent to diffuse scattering ratio. $a_i$ is the generated sample, $X$ and $Y$ are the independent and identically distributed samples (i.i.d) from the Normal distribution having zero mean and variance of 1. $Z$ is the sample from the Gamma distribution with scale parameter of 1 and shape parameter of $\alpha$. This equation generates i.i.d samples from the HK-distribution, which differs from real envelope samples of experimental echo signals due to the correlation among samples caused by the resolution cell of ultrasound (US). In order to investigate the effect of correlation on the performance of the methods, we need to generate correlated samples from the HK-distribution, which is not straightforward. In this paper, we proposed to generate correlated samples by employing correlated normal distributions of $X$ and $Y$:}
	
\begin{figure}[t]
		
		\centering
		\includegraphics[width=0.4\textwidth]{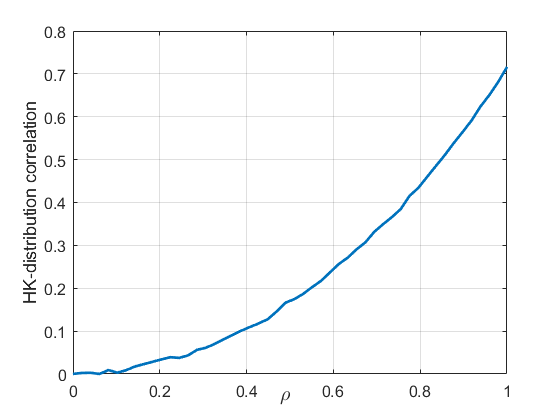}
		\caption{ The correlation coefficient of HK-distribution samples generated by Eq \ref{eq:corr} versus $\rho$ for $\alpha=3, k=0.1$.}
		\label{fig:corr}	
	\end{figure}

	\textcolor{black}{\begin{equation}
			\label{eq:corr}
			\begin{gathered}
				a_i = \sqrt{\left ( \sqrt{2k} +X_i\sigma \sqrt{Z/\alpha }\right )^2+\left (  Y_i\sigma \sqrt{Z/\alpha }\right )^2},\\
				X_i = \rho X_{i-1} + \sqrt{1-\rho^2} \mathcal{N}(0,\,1),\\
				Y_i = \rho Y_{i-1} + \sqrt{1-\rho^2} \mathcal{N}(0,\,1),
			\end{gathered}
	\end{equation}}
	\textcolor{black}{where $X_i$ and $Y_i$ are correlated with previous samples and $\rho$ controls the correlation. The correlation coefficient of the HK-distribution versus $\rho$ is illustrated in Fig. \ref{fig:corr} for $\alpha=3, k=0.1$. We selected a small value of $\rho=0.2$ for training and evaluated the performance for a higher value of $\rho$ in the Supplementary Materials.}

	\textcolor{black}{To generate training data, $log_{10}(\alpha)$ was randomly selected from values ranging from -0.3 to 1.3, corresponding to $\alpha$ of 0.5 to 20. $k$ was also randomly selected from values ranging from 0 to 1.25. As mentioned above, simulation results were reported for the sample sizes of $N_s=4096$, $N_s = 1024$, and $N_s=256$.}

	\textcolor{black}{The simulation test data was generated for 31 different values of $log_{10}(\alpha)\in\left \{-0.3,...,1.3\right \}$, and 11 values of $k\in\left \{0,...,1.25\right \}$. For each value of $log_{10}(\alpha)$ and $k$, 10 realizations were generated, giving 3410 sample sets for each sample size and $\rho$. Training and test data were generated for three sample sizes, $N_s\in\left \{4096,1024,256\right \}$, and the correlation value of $\rho\in\left \{0.2\right \}$.} \textcolor{black}{The simulation results for $\rho=0.9$ are also provided in the Supplementary Materials.} \newline{}
	\textbf{Evaluation metric for simulation test data:} \textcolor{black}{The mean absolute error (MAE) and the relative root mean square error (RRMSE) were employed to evaluate the methods for simulation test data in which the ground truth is known, and they can be defined as:}
	
	\textcolor{black}{\begin{equation}
			\begin{gathered} 
				MAE = <|y-\widetilde{y}|>,\\
				RRMSE = \sqrt{\frac{<(y-\widetilde{y})^{2}>}{|y|+\gamma }}, 
			\end{gathered}
	\end{equation}}
	\textcolor{black}{where $y$ and $\widetilde{y}$ are the ground truth and estimated parameter, respectively, $<.>$ is the averaging operation, and $\gamma$ is a small non-negative value (here we used 0.05) to avoid a division by zero.  We report MAE values in the manuscript, and RRMSE are reported in the Supplementary Materials.}    
	
	\subsubsection{Experimental Phantom Data}
	\textcolor{black}{A three-layered phantom having different scatterer number density was constructed from an emulsion of ultrafiltered milk and water-based gelatin. 5-43 $\mu m$ diameter glass beads (3000E, Potters Industries, Valley Forge, PA,  USA) were used as the source of scattering. Data were acquired with a 18L6 linear array transducer operating at a center frequency of 8.9 MHz, using a Siemens Acuson S2000 scanner (Siemens Medical Solutions USA, Inc.). Data collected from this phantom were reported in \cite{nam2012comparison}.} 

\begin{figure}[t]	
		\centering
		\includegraphics[width=0.32\textwidth]{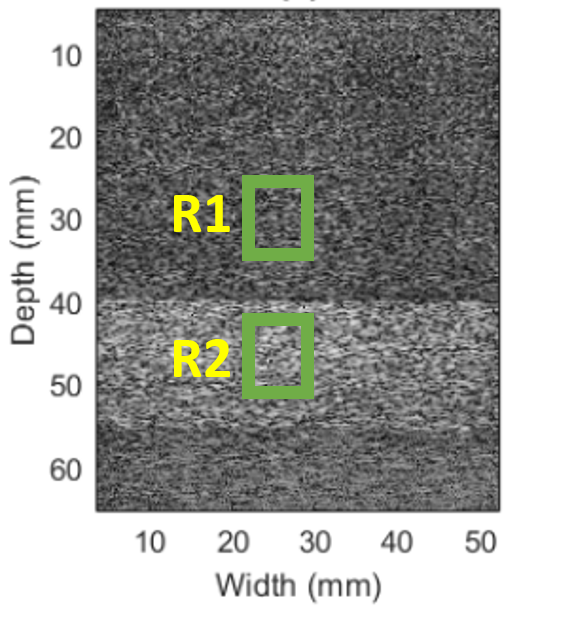}
		\caption{ \textcolor{black}{The B-mode image of the layered phantom. Two patches for extraction of the statistical features are specified. The patches are moved laterally across several frames to extract multiple features.}}
		\label{fig:bmode_exp}	
	\end{figure}

	\textcolor{black}{The B-mode image of the phantom is illustrated in Fig. \ref{fig:bmode_exp}. The top and bottom layers have the same scatterer concentration of 2 $g/L$, whereas the middle layer has a scatterer concentration of 8 $g/L$. The backscattering coefficient of the middle layer is $6.37 \times 10^{-3}$ $cm^{-1} sr^{-1}$, whereas for the top and bottom layers it is $3.52 \times 10^{-3}$ $cm^{-1} sr^{-1}$ at the center frequency of the transducer. In this phantom, the only source of intensity change to have a different backscattering coefficient is the number density of scatterers (after attenuation compensation assuming a random spatial distribution of scatterers). \textcolor{black}{According to Insana and Brown \cite{insana2022acoustic}, under the Rayleigh approximation, the backscatter coefficient is proportional to the number of scatterers within the resolution cell. Assuming that alpha is proportional to the number of scatterers \cite{insana2022acoustic}, the ratio of the values of alpha in the two regions of the phantom would be equal to the ratio of the backscatter coefficients in the same regions.} Therefore, the ratio of backscattering coefficients, $\frac{6.37 \times 10^{-3} cm^{-1} sr^{-1}}{3.52 \times 10^{-3} cm^{-1} sr^{-1}}=1.81$, can be an indicator of the true ratio of $\alpha$ which we employ as a metric to verify the estimated $\alpha$.}
	
	\textcolor{black}{A $12.6\times 7.8$ $mm$ patch size was employed for statistical feature extraction. To reduce the correlation, 5 samples in the axial direction corresponding to 1 pulse length and 1 sample in the lateral direction were skipped, which gives 3400 samples ($N_s$). The patch was moved laterally in 12 frames to obtain 60 samples of feature statistics for each of the two regions (R1 and R2). Although the exact value of $\alpha$ is unknown, the backscattering coefficient ratio can be used to verify the results.}

	\subsubsection{\textit{In vivo} Data}    
	Duck liver data were used for comparing the performance of the evaluated methods. The protocol was approved by the animal ethical care committee of  the  University  of  Montreal hospital research centre, Montreal, QC, Canada. The ducks' liver data were collected before and 14 days after force feeding to study the formation of fatty liver. Data acquisition, \textcolor{black}{and annotation of the livers} was performed as part of a study conducted by Bhatt \textit{et al.} \cite{bhatt2021multiparametric}. A Verasonics Vantage programmable system (Verasonics Inc., Kirkland, WA) with an ATL L7-4 linear probe (Philips, Bothell, WA) operating at the center frequency of 5 MHz was employed for data acquisition. Three ducks among nine were available after force feeding. The parametric images of two ducks are provided in the manuscript and the third one is given in the Supplementary Materials.    
	
	To compute statistical features, we selected patches of size $5.5$ $mm$ $\times$ $5.5$ $mm$, \textcolor{black}{and skip one sample in axial direction to reduce the correlation,} which provides $N_s=3045$ samples for each patch. These patches were moved with a $63\%$ overlap to ensure complete coverage of the liver area. \textcolor{black}{We selected smaller patch size than the one used for the experimental phantom to preserve the spatial variability of the calculated parametric images.} To minimize the impact of sample heterogeneity, we employed a patchless deep neural network, previously developed for scatterer number density regression \cite{Tehrani2023}, and applied a k-medoid clustering algorithm to select only samples that belong to the same class as the center sample. This allowed us to compute feature statistics more accurately and avoid errors arising from sample heterogeneity. \textcolor{black}{A similar strategy was adopted in \cite{roy2018assessment} where an unsupervised method was used to define image pixels into a maximum of three labels (or patches) prior to HK parameter estimation.} \textcolor{black}{An example of output of our pre-processing step is provided in the Supplementary Materials.}   
	
	\begin{figure*}[t]	
		\centering
		\includegraphics[width=0.98\textwidth]{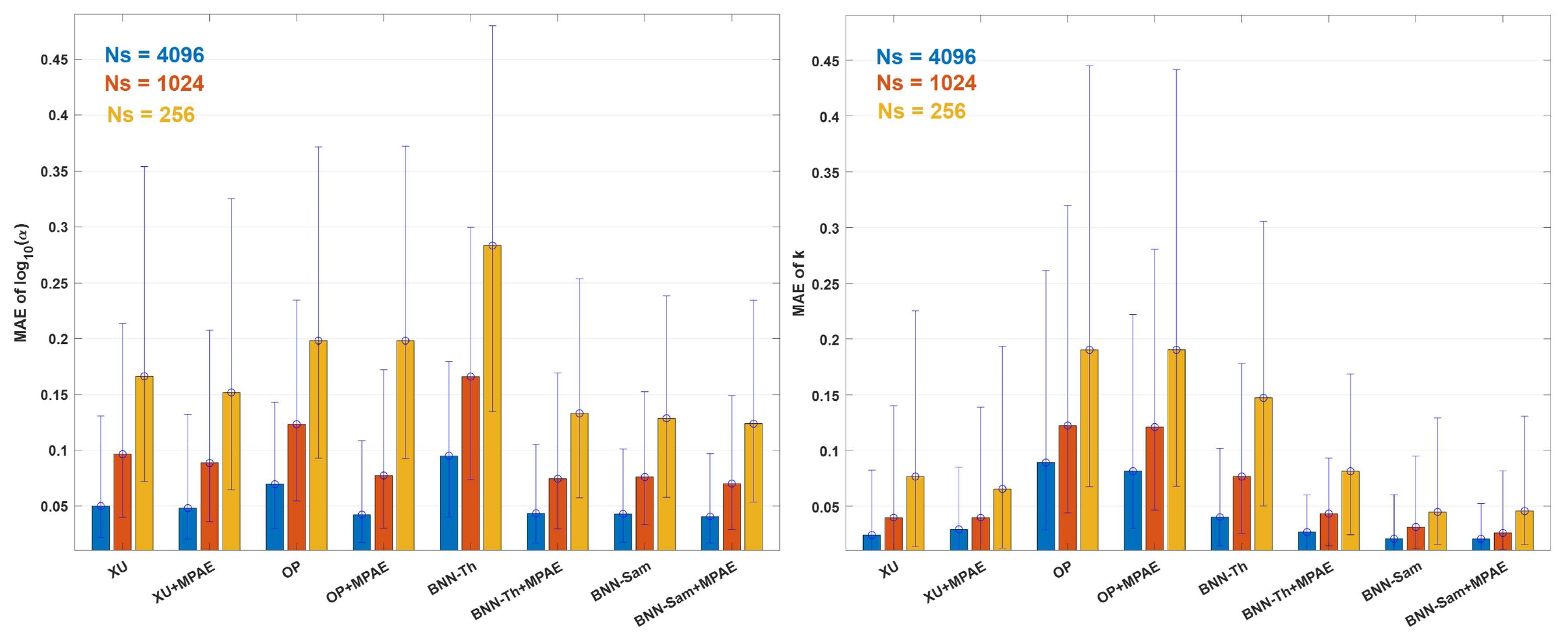}
		\caption{The median (bar height) and interquartile range (whiskers) of MAE of $log_{10}(\alpha)$ (left) and $k$ (right) for different simulated sample sizes (bar colors) and $\rho=0.2$.}
		\label{fig:sim_boxplot}	
	\end{figure*}

	\begin{figure}[t]
		
		\centering
		\includegraphics[width=0.499\textwidth]{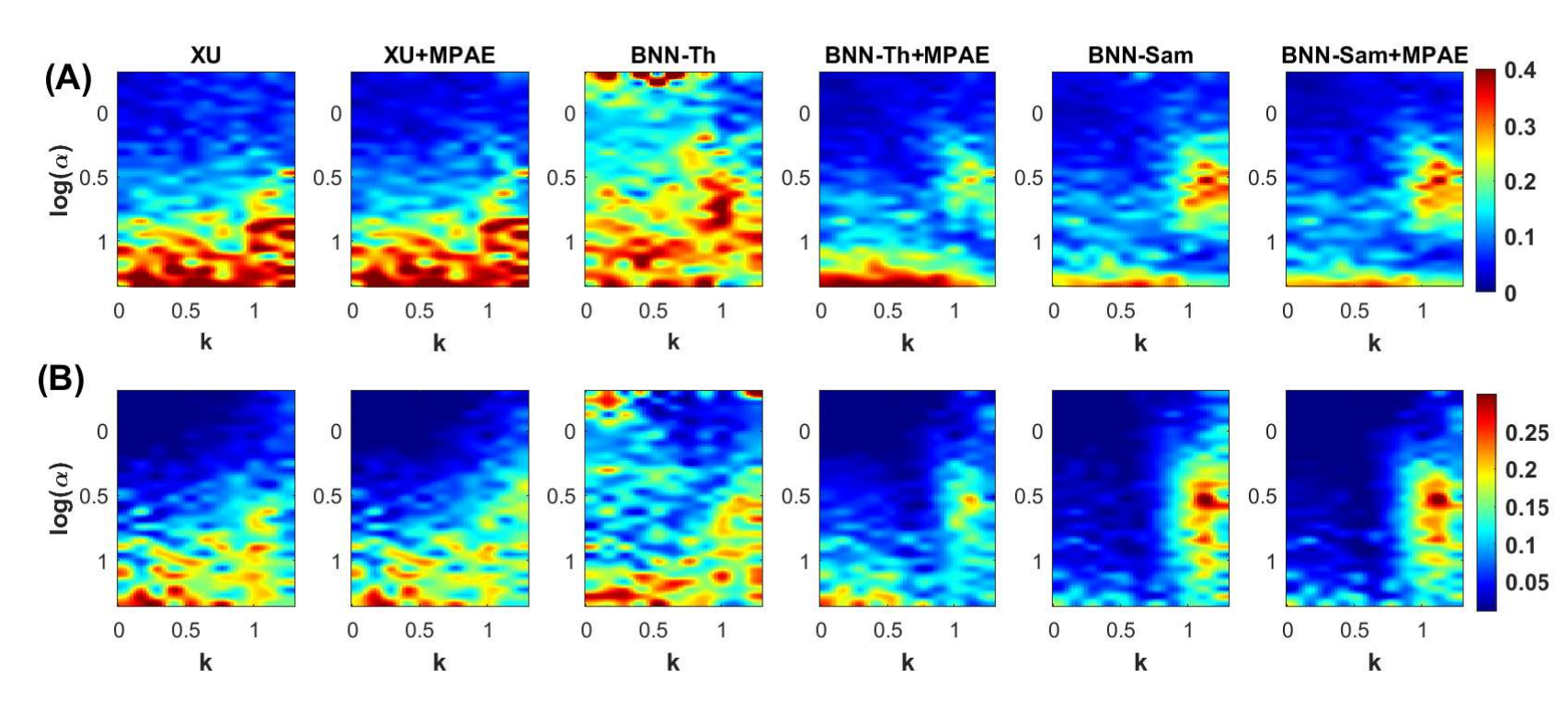}
		\caption{ The MAE error map of $log_{10}(\alpha)$ (A), and $k$ (B) for $N_s=1024,\rho=0.2$. The error is averaged over 10 realizations for each grid point. The colorbars in (A) and (B) refers to MAE of $log_{10}(\alpha)$ and $k$, respectively.}
		\label{fig:sim_map}	
	\end{figure}
	\begin{figure}[t]
		
		\centering
		\includegraphics[width=0.499\textwidth]{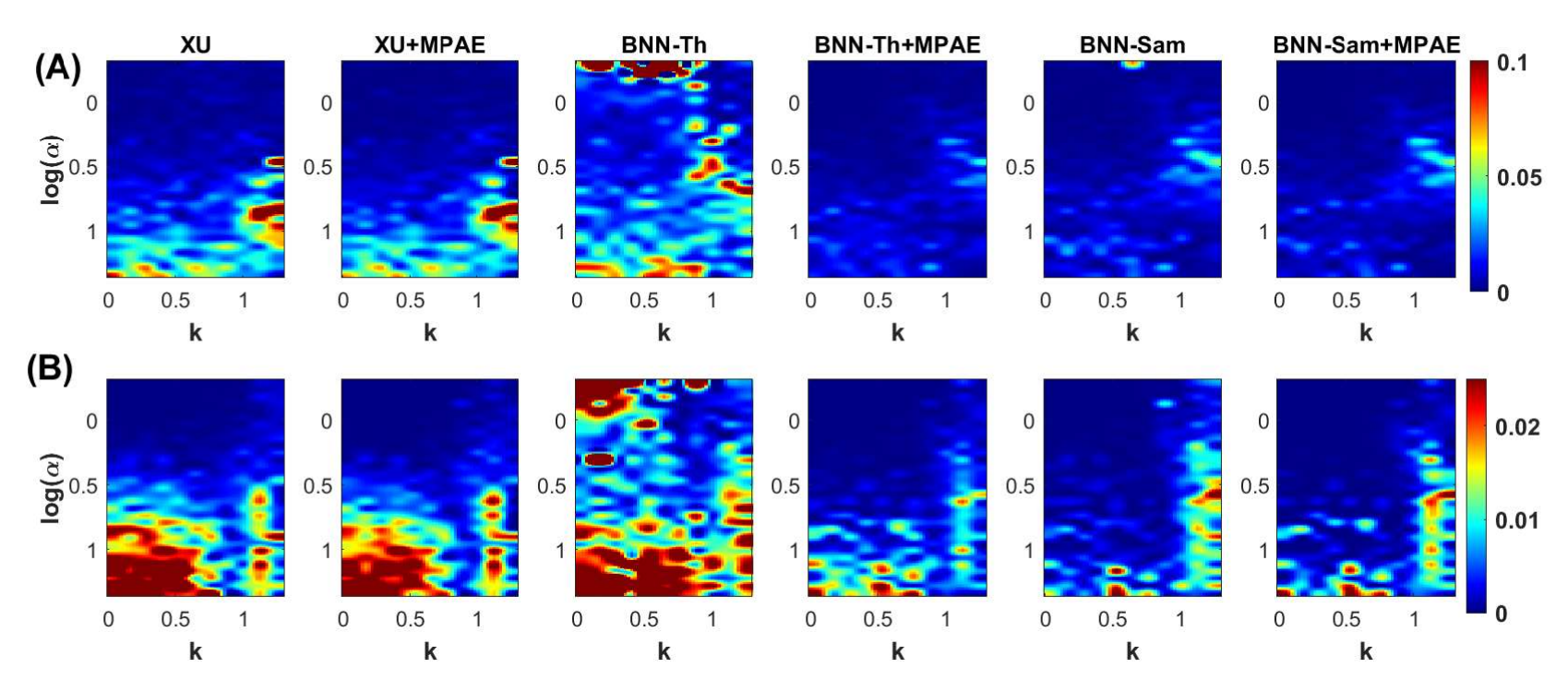}
		\caption{ The variance of MAE error map of $log_{10}(\alpha)$ (A), and $k$ (B) for $N_s=1024,\rho=0.2$. The variance is calculated over 10 realizations for each grid point. The colorbars in (A) and (B) refers to MAE of $log_{10}(\alpha)$ and $k$, respectively.}
		\label{fig:sim_map_var}	
	\end{figure}
	
	
	\begin{table}[]
		\caption{\textcolor{black}{\textit{p}-values of Wilcoxon sign tests for $log_{10}(\alpha)$. Only pairs with \textit{p}-values$>0.001$ are reported.}}
		\label{tab:pvalue_alpha}
		\resizebox{0.45\textwidth}{!}{
			\begin{tabular}{@{}ccc@{}}
				\toprule
				Pairs & Sample size & \textit{p}-values \\ \midrule
				XU, XU+MPAE & 4096 & 0.004 \\
				BNN-Sam+MPAE, BNN-Sam & 4096 & 0.052 \\
				BNN-Th+MPAE, BNN-Sam & 1024 & 0.034 \\ \bottomrule
		\end{tabular}}
	\end{table}	
	
	\begin{table}[]
		\caption{\textcolor{black}{\textit{p}-values of Wilcoxon sign tests for $k$. Only pairs with \textit{p}-values$>0.001$ are reported.}}
		\label{tab:pvalue_k}
		\resizebox{0.45\textwidth}{!}{
			\begin{tabular}{@{}ccc@{}}
				\toprule
				Pairs & Sample size & \textit{p}-values \\ \midrule
				XU, XU+MPAE & 1024 & 0.065 \\
				OP, OP+MPAE & 4096 & 0.682 \\
				BNN-Sam+MPAE, BNN-Sam & 256 & 0.158 \\
				BNN-Sam+MPAE, BNN-Sam & 4096 & 0.080 \\ \bottomrule
		\end{tabular}}
	\end{table}

		\section{Results}
		In this section we evaluate the performance of the proposed method and compare it with the XU method and the BNN.  
		\subsection{Simulation Results}
		\label{sec:sim}
		The following methods are evaluated for the simulation test data:
		
		\begin{itemize}				
			\item \textcolor{black}{XU: The method proposed by Destrempes \textit{et al.} \cite{destrempes2013estimation}.}
			\item \textcolor{black}{XU+MPAE: The method proposed by Destrempes \textit{et al.} \cite{destrempes2013estimation} but the $X$ and $U$ are obtained from denoised features of MPAE.}
			\item \textcolor{black}{OP: The method proposed by Liu \textit{et al.} \cite{liu2023study}. Only the 8 sample estimate of the features used in this manuscript are employed.} 
			\item \textcolor{black}{OP+MPAE: The method proposed by Liu \textit{et al.} \cite{liu2023study} but the denoised features of MPAE are employed.}
			\item \textcolor{black}{BNN-The: The BNN estimator is trained using the theoretical values of statistical features (i.e., solving Eq.  (4) ).}
			\item \textcolor{black}{BNN-MPAE: The BNN estimator is trained using the theoretical value of the features but testing is done using MPAE-denoised statistical features.}
			\item \textcolor{black}{BNN-Sam: The BNN estimator trained and tested using sample estimate (for each sample size, a different BNN is trained using sample estimates).}
			\item \textcolor{black}{BNN-Sam+MPAE: The BNN estimator is trained and tested using the reconstructed features from MPAE.}			
		\end{itemize}

		The median and interquartile ranges across all simulated $log_{10}(\alpha)$ and $k$ of MAE of simulation test data are illustrated in Fig. \ref{fig:sim_boxplot}. The most important observations are the following: \newline{}        
		\subsubsection{MPAE performance} \textcolor{black}{By inspecting Fig. \ref{fig:sim_boxplot} we can see that \textcolor{black}{the benefit of MPAE in denoising varies among estimators and between $\alpha$ and $k$.} To provide more details, by using MPAE there are $4\%$ ($N_s=4096$) - $8.4\%$ ($N_s=256$) reduction of median MAE of $log_{10}(\alpha)$ for the XU estimator. When using OP as the estimator, we can see $0\%$ ($N_s=256$) - $40\%$ ($N_s=4096$) reduction of median MAE of $log_{10}(\alpha)$. By using BNN-Th estimator, $53\%$ ($N_s=256$) - $55\%$ ($N_s=4096$) reduction can be achieved. On the contrary, when using BNN-Sam, the improvements are more modest compared to BNN-Th, $3.9\%$ ($N_s=256$) - $7.8\%$ ($N_s=1024$). 
			The reduction range of error for estimating $k$ after using MPAE are $-20\%$ ($N_s=4096$) - $14.4\%$ ($N_s=256$) for XU, $0\%$ ($N_s=256$) - $10\%$ ($N_s=4096$) for OP, $25\%$ ($N_s=4096$) - $44\%$ ($N_s=256$) for BNN-Th, and $0\%$ ($N_s=4096$) - $16\%$ ($N_s=256$) for BNN-Sam.}
		\newline{}
		\subsubsection{Error map} Figure \ref{fig:sim_map} shows the average of MAE in a color scale for different values of $log_{10}(\alpha)$ and $k$ for $N_s=1024$. MPAE substantially decreases BNN-Th's error, while the error maps of BNN-Sam and BNN-Sam+MPAE look similar.
		\textcolor{black}{Two estimators, namely BNN-Sam and BNN-Sam+MPAE, exhibit substantial errors in the region around $log_{10}(\alpha) = 0.5$ and $k>1$. In contrast, BNN-Th+MPAE displays a relative robustness in this region but exhibits high errors for $log_{10}(\alpha) > 1.1$. These high errors in $log_{10}(\alpha) = 0.5$ and $k>1$ is also discernible in the error visualization depicted in \cite{tehrani2022homodyned}. These errors might be attributed to the intrinsic nature of the problem, and further investigation is required.}
		\textcolor{black}{The variance maps of MAE in a color scale for different values of $log_{10}(\alpha)$ and $k$ for $N_s=1024$ are visualized in Fig. \ref{fig:sim_map_var}. BNN-Th+MPAE, BNN-Sam, and BNN-Sam+MPAE show substantially lower variance compared to XU, XU+MPAE, and BNN-Th.}\newline{}
		\subsubsection{Statistical test} A Wilcoxon sign test was performed to evaluate if the difference between the median of the errors of the methods is statistically significant. Wilcoxon sign test is a paired test that is chosen since the MAE errors are not normally distributed \cite{mcdonald2009handbook}. The \textit{p}-values of the pairs larger than 0.001 are reported in Table \ref{tab:pvalue_alpha} for $log_{10}(\alpha)$ and in Table \ref{tab:pvalue_k} for $k$. Only a few pairs reported in the tables were not statistically different.
		
		\subsubsection{Investigating different BNNs}
		By inspecting the results of different BNNs in Fig. \ref{fig:sim_boxplot}, we can find that:
		\newline{}
		\textcolor{black}{\textbf{Training using sample estimates:} The BNN trained by sample estimates of the features (BNN-Sam) outperforms the one trained by theoretical values (BNN-Th). The reason behind this improvement is that in contrast to BNN-Th, in BNN-Sam, for each specific sample size an individual BNN is trained and the network learns the distribution of the noise as well. It is also comparable to BNN+MPAE which shows that MPAE can act as the denoiser prior to the BNN-Th estimator}\newline{}
		\textcolor{black}{\textbf{Training using reconstructed features:} It can be observed that the BNN trained on reconstructed features (BNN-sam+MPAE) outperforms BNN+MPAE which implies that the BNN is able to learn the noise distribution of the reconstructed features of MPAE. This method also slightly outperform BNN-Sam. The median values of the MAE errors of $log_{10}(\alpha)$ of BNN-Sam+MPAE are 0.123, 0.070, and 0.041 for sample sizes 256, 1024, and 4096, respectively. While, they are 0.128, 0.076, and 0.043 for BNN-Sam. For $k$, the MAE errors of BNN-Sam+MPAE are 0.046, 0.026, and 0.020. While BNN-Sam achieved the errors of 0.045, 0.031, and 0.020.}\newline{}
		\subsubsection{MPAE bottleneck size} \textcolor{black}{We employed a bottleneck size of 3 for the reported results in the paper. Table \ref{tab:bott} shows the MAE error for BNN-Th+MPAE using bottleneck sizes of two and three. We did not use higher dimensions of bottleneck size since there are 8 features for which three of them are highly correlated with the other three features ($[ R^{0.72}, S^{0.72}, K^{0.72}]$ with $[ R^{0.88}, S^{0.88}, K^{0.88}]$); therefore, there are 5 features that are reduced to a three-feature space. We obtained lower error using a bottleneck size of three compared to the size of two; hence, we used that throughout the paper.}
		\begin{table}[]
			\caption{\textcolor{black}{Ablation experiment of the bottleneck size of MPAE. Results are reported by median \small[$25\%$,$75\%$].}}
			\label{tab:bott}
			\resizebox{0.495\textwidth}{!}{
				\begin{tabular}{@{}ccc|cc@{}}
					\toprule
					& \multicolumn{2}{c|}{$log_{10}(\alpha)$} & \multicolumn{2}{c}{$k$} \\ \midrule
					$N_s$	& MPAE-b2 & MPAE-b3 & MPAE-b2 & MPAE-b3 \\ \midrule
					4096 & 0.066 {[}0.025,0.168{]} & 0.041 {[}0.017,0.097{]} & 0.146 {[}0.047,0.303{]} & 0.020 {[}0.007,0.053{]} \\
					1024 & 0.084 {[}0.034,0.190{]} & 0.070 {[}0.028,0.149{]} & 0.154 {[}0.058,0.309{]} & 0.026 {[}0.010,0.081{]} \\
					256 & 0.136 {[}0.058,0.276{]} & 0.123 {[}0.054,0.234{]} & 0.199 {[}0.080,0.382{]} & 0.047 {[}0.015,0.130{]} \\ \bottomrule
			\end{tabular}}
		\end{table}

		\subsubsection{Effect of Correlation} \textcolor{black}{The results given in Fig. \ref{fig:sim_boxplot} are with correlation ($\rho$) of 0.2 (both training and test data). To investigate the effect of correlation on the estimated parameters, we evaluated the proposed methods on high correlation of 0.9 (networks are trained with $\rho=0.2$) in the Supplementary Materials. \textcolor{black}{The results show a slight increase of the error of the estimators for the high correlation of 0.9.}}

		\begin{figure}[t]	
			\centering
			\includegraphics[width=0.499\textwidth]{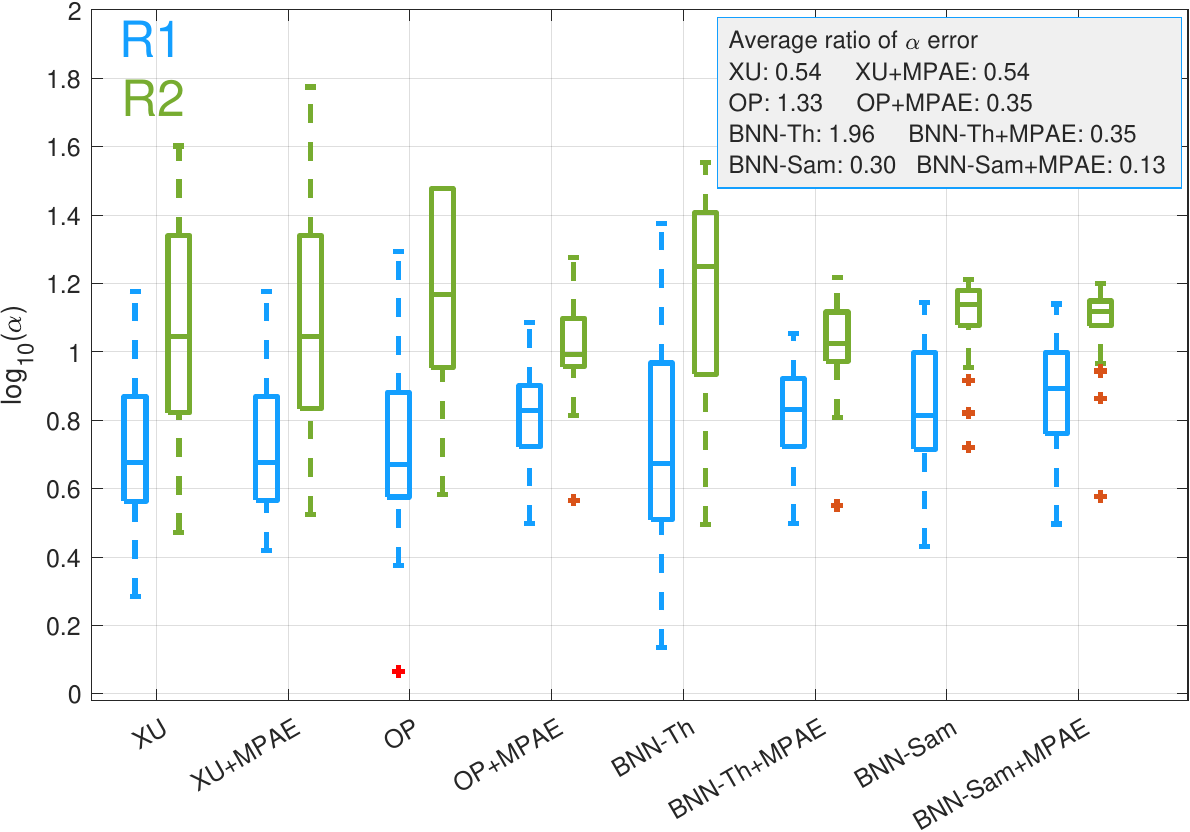}
			\caption{ \textcolor{black}{The boxplot of $log_{10}(\alpha)$ of the layered phantom using the evaluated methods. Most methods achieve closer ratio to the correct one (1.81) after using MPAE (except XU+MPAE). Outliers (values more than 1.5 times of interquartile range away from the bottom and top of the box) are shown as red $"+"$.} }
			\label{fig:exp_boxplot}	
		\end{figure}
		\subsection{Experimental Phantom Results}
		\label{sec:phantom}
		\textcolor{black}{The box plot of $\alpha$ of the evaluated methods for regions R1 and R2 (specified in Fig. \ref{fig:bmode_exp}) is illustrated in Fig. \ref{fig:exp_boxplot}, the box plot of $k$ and \textcolor{black}{the \textit{p}-values of statistical test} are also provided in the Supplementary Materials. The assumed ratio of $\alpha$ for R1 and R2 is 1.81 (obtained from the backscattering coefficient). The median values of the ratios of the estimated $\alpha$ are 2.345, 2.345, 3.140, 1.463, 3.770, 1.563, 2.110, and 1.682 for XU, XU+MPAE, OP, OP+MPAE, BNN-Th, BNN-Th+MPAE, BNN-Sam, and BNN-Sam+MPAE, respectively. Most methods achieve closer ratio to the correct one (1.81) after using MPAE (except XU+MPAE which gave the same ratio as XU).}
		\begin{figure*}[t]	
			\centering
			\captionsetup{justification=centering}
			\includegraphics[width=0.7\textwidth]{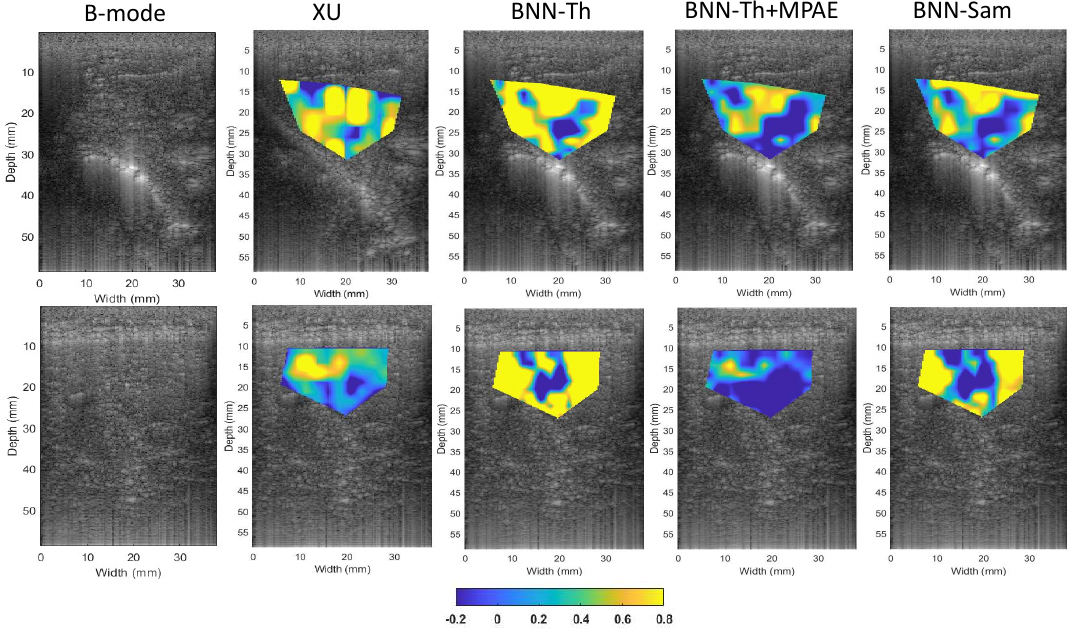}
			\centering
			\caption{\centering The B-mode image (first column) and the parametric image of $log_{10}(\alpha)$ of duck A before force feeding (top row), and after (bottom row). }
			\label{fig:DA_T0_param}	
		\end{figure*}
		
		\begin{figure*}[t]	
			\centering
			\captionsetup{justification=centering}
			\includegraphics[width=0.7\textwidth]{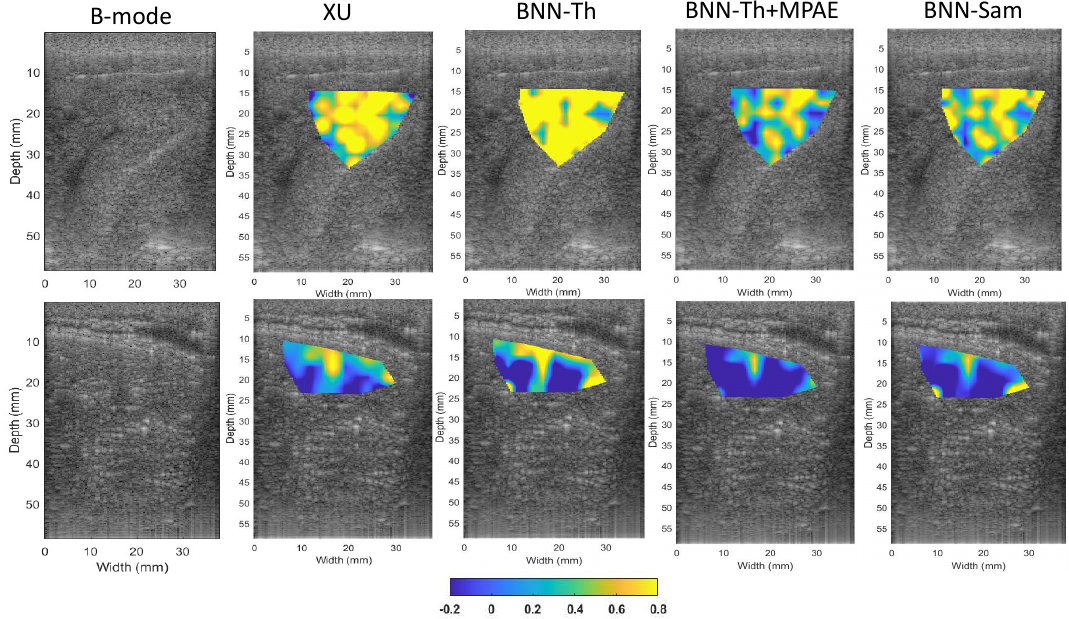}
			\centering
			\caption{\centering The B-mode image (first column) and the parametric image of $log_{10}(\alpha)$ of duck B before force feeding (top row), and after (bottom row). }
			\label{fig:DB_T0_param}	
		\end{figure*}

		\begin{table}[t]
			\caption{The correlation between the parametric images of XU and the other evaluated methods.}
			\label{tab:corr}
			\resizebox{0.499\textwidth}{!}{
				\begin{tabular}{@{}ccccc@{}}
					\toprule
					Methods & \begin{tabular}[c]{@{}c@{}}Duck A\\ before force feeding\end{tabular} & \begin{tabular}[c]{@{}c@{}}Duck B\\ before force feeding\end{tabular} & \begin{tabular}[c]{@{}c@{}}Duck A\\ after force feeding\end{tabular} & \begin{tabular}[c]{@{}c@{}}Duck B\\ after force feeding\end{tabular} \\ \midrule
					XU, BNN-Th & 0.47 & 0.05 & 0.47 & 0.50 \\
					XU, BNN-Sam & 0.40 & 0.04 & 0.52 & 0.50 \\
					XU, BNN-Th+MPAE & \textbf{0.67} & \textbf{0.81} & \textbf{0.82} & \textbf{0.78} \\ \bottomrule
			\end{tabular}}
		\end{table}

		\begin{figure}[t]	
			\centering
			\includegraphics[width=0.47\textwidth]{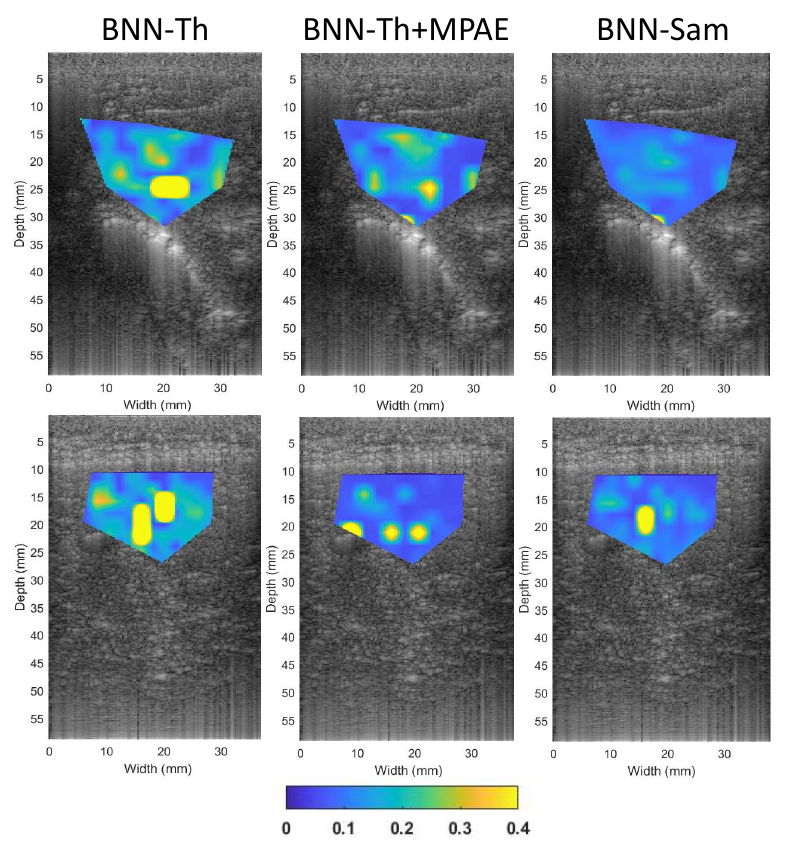}
			\caption{ The parametric image of uncertainty of BNN of duck A before force feeding (top row), and after (bottom row). }
			\label{fig:DA_T0_param_unc}	
		\end{figure}
		\begin{figure}[]	
			\centering
			\includegraphics[width=0.47\textwidth]{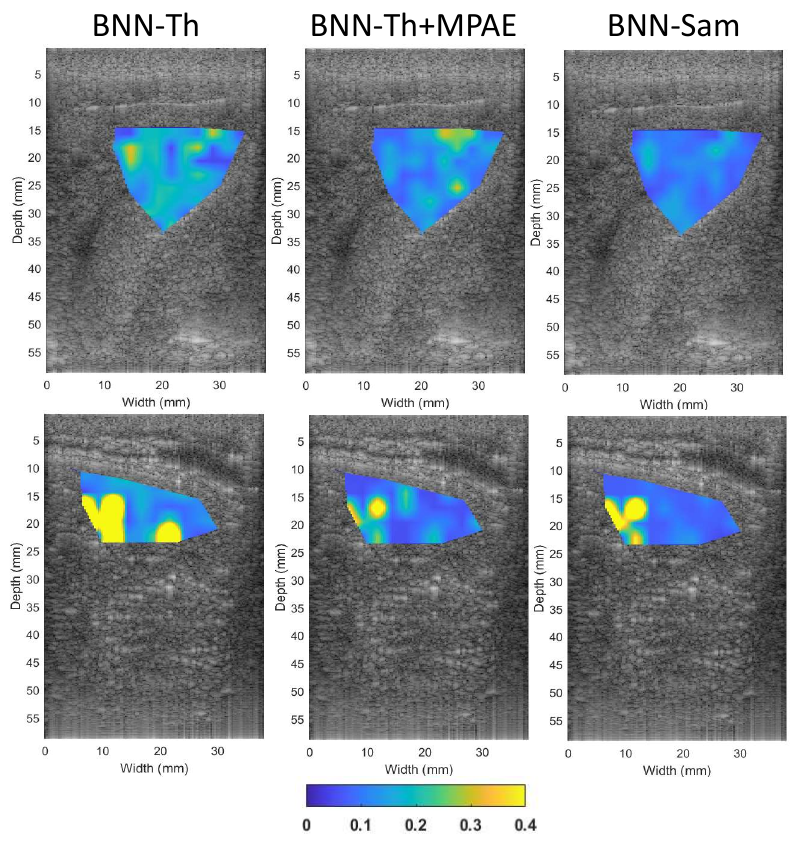}
			\caption{ The parametric image of uncertainty of BNN of duck B before force feeding (top row), and after (bottom row). }
			\label{fig:DB_T0_param_unc}	
		\end{figure}
		\textcolor{black}{The variances of estimated $log_{10}(\alpha)$ of the R1 region are 0.044, 0.041, 0.058, 0.017, 0.079, 0.018, 0.030, and 0.023 for XU, XU+MPAE, OP, OP+MPAE, BNN-Th, BNN-Th+MPAE, BNN-Sam, and BNN-Sam+MPAE, respectively, and 0.097, 0.120, 0.069, 0.013, 0.076, 0.013, 0.009, and 0.009 for region R2. It can be observed that by adding MPAE to the estimators, the variance is decreased in most cases. The over-estimation of $\alpha$ and the high variance can be observed in XU and BNN-Th results. Zhou \textit{et al.} showed that when the sample size is not large enough, an over-estimation of $\alpha$ occurs \cite{zhou2020value}, which is also confirmed by our results. BNN-Sam+MPAE has the closest ratio of $\alpha$ values (1.682) to the actual ratio (1.81).}

		%

		\subsection{\textit{In vivo} Results}
		\label{sec:invivo}
		\textcolor{black}{The parametric images of $log_{10}(\alpha)$ are illustrated in Fig. \ref{fig:DA_T0_param} for duck A and in Fig. \ref{fig:DB_T0_param} for duck B before force feeding (top row) and after (bottom row), and their parametric images of $k$ are provided in the Supplementary Materials. Gesnik \textit{et al.} \cite{gesnik2020vivo} reported that the average of $1/\alpha $ of the whole liver region considering all studied ducks increases from $1/\alpha=0.69 \pm 0.10$ to $1/\alpha=0.94 \pm 0.07$ after 14 days of force feeding. Therefore, generally, lower $\alpha$ values are expected after force feeding.} \textcolor{black}{According to the given parametric images, XU and BNN-Th+MPAE results present similar spatial variability of the quantitative features. XU and BNN-Th+MPAE both produce high values in the same locations; while BNN-Th produces very high values of $log_{10}(\alpha)$ in most parts. BNN-Sam also provided close results to BNN-Th+MPAE. The results of BNN-Sam+MPAE were not provided since we observed that they are very similar to BNN-Sam.} 
		
		The correlation between XU and other estimators are reported in Table \ref{tab:corr}, which indicates higher agreement between BNN-Th+MPAE and XU than the other estimators with XU. Both ducks' liver $log_{10}(\alpha)$ parametric images obtained by XU and BNN-Th+MPAE have reduction of portions with high values after force feeding (Figs. \ref{fig:DA_T0_param} and \ref{fig:DB_T0_param}), which is expected as a spatially averaged reduction was reported in \cite{gesnik2020vivo}. The box plots of the estimated $log_{10}(\alpha)$ for the three ducks are also provided in the Supplementary Materials.

		\textcolor{black}{Uncertainty of the estimation is another aspect that can be investigated. As discussed earlier, the BNN estimator can also provide the uncertainty by taking the variance of the estimates using different realization of the network's weights. The parametric images of the uncertainty of the BNN-Th, BNN-Th+MPAE, and BNN-Sam are illustrated in Fig. \ref{fig:DA_T0_param_unc} for duck A, and in Fig. \ref{fig:DB_T0_param_unc} for duck B. It is clear that BNN-Sam and BNN-Th+MPAE have substantially lower uncertainty compared to BNN-Th.}  
		\section{Discussion and Future Work}
		\textcolor{black}{In this paper, we presented a model projection autoencoder to reconstruct clean statistical features used in the estimation of HK parameters from the amplitude of the detected echo signals from noisy sample estimates. We also investigated different BNN estimators' training approaches and how to pair them with the autoencoder. Any HK parameter estimator can be used to estimate HK parameters from MPAE-denoised features. 
			Source of the noise is an aspect that should be investigated further. In this paper, the noise emanates from the low sample size. Other sources of noise including the presence of outlier samples or the accuracy of the HK model of the true echo amplitude distribution within the patch requires further investigation.}
		
		\textcolor{black}{We observed more substantial improvement by employing MPAE for the estimators that are blind to the noise distribution (XU,OP, and BNN-Th) compared to the one that has learned the noise distribution (BNN-Sam). The main reason is that when the BNN is trained on the sample estimate of the features (BNN-Sam), it also learns the noise distribution of that specific sample size; therefore, it becomes more robust to the noisy features.}
		
		\textcolor{black}{The simulation test results indicate that the enhancements achieved by utilization of MPAE are less noticeable for the parameter $k$ compared to $\alpha$. This observation could be attributed to the features employed. Incorporating additional features, such as $\Omega$ (the parameter of the Nakagami distribution), might address this issue. Another way that might solve this issue is to train separate BNNs for $\alpha$ and $k$.}
		
		\textcolor{black}{We should note that by observing Fig. \ref{fig:sim_map}, it is noticeable that BNN-based methods exhibit higher error in the region of $log_{10}(\alpha)\approx0.5$ and $k>1$; while XU method has a high error in $log_{10}(\alpha)>1$. This shows that optimum estimator may vary depending on the value of $\alpha$ and $k$. We also checked our implementation to verify the correctness of the implementation of theoretical feature values. We
			generated 20000 random samples drawn from the HK-distribution to ensure the effect of lack of samples is minimized. In the next step, we calculated sample
			estimates of statistical features from the generated samples, and compared them with the
			theoretical value. We found a very low and negligible value of error (less than 0.001) which shows that the implementation is
			correct.}
		
		\textcolor{black}{Another interesting area to investigate the effect of sample/patch size is through
			beam steering, where additional samples become available by insonifying the ROI
			using a different beam direction \cite{rivaz20079c}.} 
		
		\textcolor{black}{In Section \ref{sec:phantom}, the validation of the experimental findings was conducted through comparing with the backscattering coefficient ratio. However, it is important to note that this validation is contingent upon certain conditions or hypotheses being satisfied, namely the low ratio of coherent to diffuse scattering, a spatially uniform random distribution of scatterers within the patch, and proper compensation for total attenuation \cite{destrempes2021statistical,destrempes2015unifying,cloutier2021quantitative}.}
		
		\textcolor{black}{In Section \ref{sec:invivo}, the parametric images of $log_{10}(\alpha)$ were investigated and displayed. The box plots of $log_{10}(\alpha)$ values of the three ducks were provided in the Supplementary Materials which showed that separability of the $log_{10}(\alpha)$ was improved by employing MPAE. We selected a patch size of $5.5$ $mm$ $\times$ $5.5$ $mm$, which provided $N_s=3045$ samples for estimating the statistical features. We noticed noisier (higher variance) parametric images when smaller patch size was employed. By increasing the size of the patch, the parametric image looked smoother but the spatial resolution was reduced. The selected patch size resulted in a good balance between the variance and the spatial resolution.}

\begin{figure}[t]	
			\centering
			\includegraphics[width=0.45\textwidth]{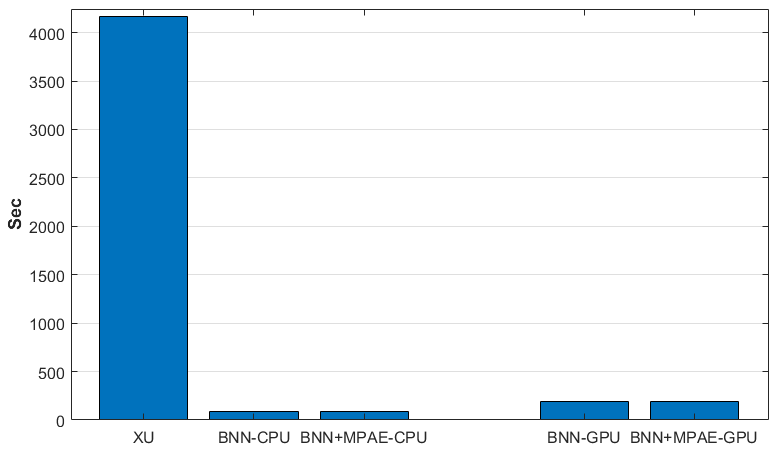}
			\caption{ \textcolor{black}{The processing time for 500 patches. The XU was implemented on MATLAB and executed on CPU. For the batch size of 1, the processing time of BNN and BNN+MPAE is lower on CPU compared to GPU.} }
			\label{fig:time}	
		\end{figure}

		\textcolor{black}{Both the BNN and MPAE were implemented using PyTorch and were trained on a single Nvidia RTX 3090 GPU. MPAE requires $0.57\pm0.049$ ms for processing, whereas BNN takes $7.5\pm0.3$ ms for each inference on GPU. For the batch size of 1 (which is the batch size in our implementation), the processing time on CPU is lower compared to GPU; achieving $0.33\pm0.032$ and $3.6\pm0.29$ for MPAE and BNN, respectively. In order to estimate parameters and quantify the uncertainty, multiple inferences are required; therefore, we executed the BNN 50 times per observation, resulting in a total processing time of $7.5\pm0.3$ ms $\times$ 50 = $377\pm15.4$ ms on GPU. It should be noted that the processing times are reported for the batch size of 1; GPU performance is expected to be improved and outperform CPU performance if a higher batch size is employed. No efforts were made to optimize the code which can be a topic of feature research. The XU and OP estimators were implemented on MATLAB and executed on an i7 CPU, and the processing time was varied with an average and standard deviation of $8.33\pm7.26$ and $19.50\pm4.8$ sec, respectively, for each patch.} \textcolor{black}{The processing time is an important factor in HK parametric image calculation since there might be hundreds of patches within the region of interest. For instance, for a ROI containing 500 patches, XU estimator takes an average of 69.4 minutes. Whereas, BNN only takes 3.14 minutes on average when executed 50 times per observation (The number of executions can be reduced to accelerate the computation), which is 16 times faster than XU. The processing times for 500 patches are illustrated in Fig. \ref{fig:time}. It should be noted that no optimization has been done on the implementation of XU and OP. A more optimized implementation can speed up those methods.}
		
		\textcolor{black}{In this paper, we proposed the MPAE as a pre-processing stage to be used before the HK parameter estimators, and we also investigated different training approaches for the BNN estimator. One possible area of future work is to combine MPAE and the BNN. The two networks can be trained in an end-to-end fashion, which can reduce the computational complexity of the method.} 
		
		%
		%
		%
		
		\section{Conclusion}
		\textcolor{black}{In this paper, we proposed a model projection neural network based on denoising autoencoders to reconstruct refined statistical features to improve the HK-distribution parameter estimation. We also thoroughly investigated different training strategies of the BNN. The proposed methods were validated using simulation data, experimental phantom, and clinical data. }\textcolor{black}{Our findings showed that employing MPAE-denoised features can reduce the errors of estimators especially for low sample sizes. Comparing the two estimated parameters, we observed more substantial improvements in estimation of $log_{10}(\alpha)$ than $k$.} 
		\newline{}
		\section*{Acknowledgment}
		\textcolor{black}{We acknowledge funding from the Natural Sciences and Engineering Research Council of Canada (NSERC), and the Fonds de recherche du Québec-Santé (FRQS $\#$34939).} 
		\newline{}
		
		\bibliographystyle{IEEEtran}
		\bibliography{refs3}

\begin{thebibliography}{10}
\providecommand{\url}[1]{#1}
\csname url@samestyle\endcsname
\providecommand{\newblock}{\relax}
\providecommand{\bibinfo}[2]{#2}
\providecommand{\BIBentrySTDinterwordspacing}{\spaceskip=0pt\relax}
\providecommand{\BIBentryALTinterwordstretchfactor}{4}
\providecommand{\BIBentryALTinterwordspacing}{\spaceskip=\fontdimen2\font plus
\BIBentryALTinterwordstretchfactor\fontdimen3\font minus
  \fontdimen4\font\relax}
\providecommand{\BIBforeignlanguage}[2]{{%
\expandafter\ifx\csname l@#1\endcsname\relax
\typeout{** WARNING: IEEEtran.bst: No hyphenation pattern has been}%
\typeout{** loaded for the language `#1'. Using the pattern for}%
\typeout{** the default language instead.}%
\else
\language=\csname l@#1\endcsname
\fi
#2}}
\providecommand{\BIBdecl}{\relax}
\BIBdecl

\bibitem{oelze2016review}
M.~L. Oelze and J.~Mamou, ``Review of quantitative ultrasound: Envelope
  statistics and backscatter coefficient imaging and contributions to
  diagnostic ultrasound,'' \emph{IEEE transactions on ultrasonics,
  ferroelectrics, and frequency control}, vol.~63, no.~2, pp. 336--351, 2016.

\bibitem{wagner1983statistics}
R.~F. Wagner, ``Statistics of speckle in ultrasound b-scans,'' \emph{IEEE
  Trans. Sonics \& Ultrason.}, vol.~30, no.~3, pp. 156--163, 1983.

\bibitem{jafarpisheh2020analytic}
N.~Jafarpisheh, T.~J. Hall, H.~Rivaz, and I.~M. Rosado-Mendez, ``Analytic
  global regularized backscatter quantitative ultrasound,'' \emph{IEEE
  Transactions on Ultrasonics, Ferroelectrics, and Frequency Control}, 2020.

\bibitem{Vajihi2018}
Z.~{Vajihi}, I.~M. {Rosado-Mendez}, T.~J. {Hall}, and H.~{Rivaz}, ``Low
  variance estimation of backscatter quantitative ultrasound parameters using
  dynamic programming,'' \emph{IEEE Transactions on Ultrasonics,
  Ferroelectrics, and Frequency Control}, vol.~65, no.~11, pp. 2042--2053,
  2018.

\bibitem{yao1990backscatter}
L.~X. Yao, J.~A. Zagzebski, and E.~L. Madsen, ``Backscatter coefficient
  measurements using a reference phantom to extract depth-dependent
  instrumentation factors,'' \emph{Ultrasonic imaging}, vol.~12, no.~1, pp.
  58--70, 1990.

\bibitem{rouyer2016vivo}
J.~Rouyer, T.~Cueva, T.~Yamamoto, A.~Portal, and R.~J. Lavarello, ``In vivo
  estimation of attenuation and backscatter coefficients from human thyroids,''
  \emph{IEEE transactions on ultrasonics, ferroelectrics, and frequency
  control}, vol.~63, no.~9, pp. 1253--1261, 2016.

\bibitem{soylu2023calibrating}
U.~Soylu and M.~L. Oelze, ``Calibrating data mismatches in deep learning-based
  quantitative ultrasound using setting transfer functions,'' \emph{IEEE
  Transactions on Ultrasonics, Ferroelectrics, and Frequency Control}, 2023.

\bibitem{mohammadi2021ultrasound}
N.~Mohammadi, M.~M. Doyley, and M.~Cetin, ``Ultrasound elasticity imaging using
  physics-based models and learning-based plug-and-play priors,'' in
  \emph{ICASSP 2021-2021 IEEE International Conference on Acoustics, Speech and
  Signal Processing (ICASSP)}.\hskip 1em plus 0.5em minus 0.4em\relax IEEE,
  2021, pp. 1165--1169.

\bibitem{tehrani2020displacement}
A.~K. Tehrani and H.~Rivaz, ``Displacement estimation in ultrasound
  elastography using pyramidal convolutional neural network,'' \emph{IEEE
  Transactions on Ultrasonics, Ferroelectrics, and Frequency Control}, 2020.

\bibitem{yazdani2022revisited}
L.~Yazdani, M.~Bhatt, I.~Rafati, A.~Tang, and G.~Cloutier, ``The revisited
  frequency-shift method for shear wave attenuation computation and imaging,''
  \emph{IEEE Transactions on Ultrasonics, Ferroelectrics, and Frequency
  Control}, vol.~69, no.~6, pp. 2061--2074, 2022.

\bibitem{rosado2016analysis}
I.~M. Rosado-Mendez, L.~C. Drehfal, J.~A. Zagzebski, and T.~J. Hall, ``Analysis
  of coherent and diffuse scattering using a reference phantom,'' \emph{IEEE
  transactions on ultrasonics, ferroelectrics, and frequency control}, vol.~63,
  no.~9, pp. 1306--1320, 2016.

\bibitem{destrempes2013review}
F.~Destrempes and G.~Cloutier, ``Review of envelope statistics models for
  quantitative ultrasound imaging and tissue characterization,''
  \emph{Quantitative ultrasound in soft tissues}, pp. 219--274, 2013.

\bibitem{dutt1995speckle}
V.~Dutt and J.~F. Greenleaf, ``Speckle analysis using signal to noise ratios
  based on fractional order moments,'' \emph{Ultrasonic Imaging}, vol.~17,
  no.~4, pp. 251--268, 1995.

\bibitem{destrempes2015unifying}
F.~Destrempes, E.~Franceschini, T.~Fran{\c{c}}ois, and G.~Cloutier, ``Unifying
  concepts of statistical and spectral quantitative ultrasound techniques,''
  \emph{IEEE transactions on medical imaging}, vol.~35, no.~2, pp. 488--500,
  2015.

\bibitem{destrempes2022quantitative}
F.~Destrempes, M.~Gesnik, B.~Chayer, M.-H. Roy-Cardinal, D.~Olivi{\'e}, J.-M.
  Giard, G.~Sebastiani, B.~N. Nguyen, G.~Cloutier, and A.~Tang, ``Quantitative
  ultrasound, elastography, and machine learning for assessment of steatosis,
  inflammation, and fibrosis in chronic liver disease,'' \emph{Plos one},
  vol.~17, no.~1, p. e0262291, 2022.

\bibitem{fang2020ultrasound}
F.~Fang, J.~Fang, Q.~Li, D.-I. Tai, Y.-L. Wan, K.~Tamura, T.~Yamaguchi, and
  P.-H. Tsui, ``Ultrasound assessment of hepatic steatosis by using the double
  nakagami distribution: a feasibility study,'' \emph{Diagnostics}, vol.~10,
  no.~8, p. 557, 2020.

\bibitem{zhou2019hepatic}
Z.~Zhou, Q.~Zhang, W.~Wu, Y.-H. Lin, D.-I. Tai, J.-H. Tseng, Y.-R. Lin, S.~Wu,
  and P.-H. Tsui, ``Hepatic steatosis assessment using ultrasound homodyned-k
  parametric imaging: the effects of estimators,'' \emph{Quantitative Imaging
  in Medicine and Surgery}, vol.~9, no.~12, p. 1932, 2019.

\bibitem{zhou2020value}
Z.~Zhou, J.~Fang, A.~Cristea, Y.-H. Lin, Y.-W. Tsai, Y.-L. Wan, K.-M. Yeow,
  M.-C. Ho, and P.-H. Tsui, ``Value of homodyned k distribution in ultrasound
  parametric imaging of hepatic steatosis: An animal study,''
  \emph{Ultrasonics}, vol. 101, p. 106001, 2020.

\bibitem{nguyen2019reference}
T.~Nguyen and M.~Oelze, ``Reference free quantitative ultrasound classification
  of fatty liver,'' in \emph{2019 IEEE International Ultrasonics Symposium
  (IUS)}.\hskip 1em plus 0.5em minus 0.4em\relax IEEE, 2019, pp. 2424--2427.

\bibitem{muhtadi2022breast}
S.~Muhtadi, ``Breast tumor classification using intratumoral quantitative
  ultrasound descriptors,'' \emph{Computational and Mathematical Methods in
  Medicine}, vol. 2022, 2022.

\bibitem{chowdhury2022ultrasound}
A.~Chowdhury, R.~R. Razzaque, S.~Muhtadi, A.~Shafiullah, E.~U.~I. Abir, B.~S.
  Garra, and S.~K. Alam, ``Ultrasound classification of breast masses using a
  comprehensive nakagami imaging and machine learning framework,''
  \emph{Ultrasonics}, vol. 124, p. 106744, 2022.

\bibitem{destrempes2020added}
F.~Destrempes, I.~Trop, L.~Allard, B.~Chayer, J.~Garcia-Duitama, M.~El~Khoury,
  L.~Lalonde, and G.~Cloutier, ``Added value of quantitative ultrasound and
  machine learning in bi-rads 4--5 assessment of solid breast lesions,''
  \emph{Ultrasound in Medicine \& Biology}, vol.~46, no.~2, pp. 436--444, 2020.

\bibitem{trop2015added}
I.~Trop, F.~Destrempes, M.~El~Khoury, A.~Robidoux, L.~Gaboury, L.~Allard,
  B.~Chayer, and G.~Cloutier, ``The added value of statistical modeling of
  backscatter properties in the management of breast lesions at us,''
  \emph{Radiology}, vol. 275, no.~3, pp. 666--674, 2015.

\bibitem{byra2016classification}
M.~Byra, A.~Nowicki, H.~Wr{\'o}blewska-Piotrzkowska, and K.~Dobruch-Sobczak,
  ``Classification of breast lesions using segmented quantitative ultrasound
  maps of homodyned k distribution parameters,'' \emph{Medical physics},
  vol.~43, no.~10, pp. 5561--5569, 2016.

\bibitem{hoerig2023classification}
C.~Hoerig, K.~Wallace, M.~Wu, and J.~Mamou, ``Classification of metastatic
  lymph nodes in vivo using quantitative ultrasound at clinical frequencies,''
  \emph{Ultrasound in Medicine \& Biology}, vol.~49, no.~3, pp. 787--801, 2023.

\bibitem{Hruska2009}
D.~P. Hruska and M.~L. Oelze, ``Improved parameter estimates based on the
  homodyned k distribution,'' \emph{IEEE transactions on ultrasonics,
  ferroelectrics, and frequency control}, vol.~56, no.~11, pp. 2471--2481,
  2009.

\bibitem{destrempes2013estimation}
F.~Destrempes, J.~Por{\'e}e, and G.~Cloutier, ``Estimation method of the
  homodyned k-distribution based on the mean intensity and two log-moments,''
  \emph{SIAM journal on imaging sciences}, vol.~6, no.~3, pp. 1499--1530, 2013.

\bibitem{liu2023study}
Y.~Liu, B.~He, Y.~Zhang, X.~Lang, R.~Yao, and L.~Pan, ``A study on a parameter
  estimator for the homodyned k distribution based on table search for
  ultrasound tissue characterization,'' \emph{Ultrasound in Medicine \&
  Biology}, 2023.

\bibitem{zhou2021parameter}
Z.~Zhou, A.~Gao, W.~Wu, D.-I. Tai, J.-H. Tseng, S.~Wu, and P.-H. Tsui,
  ``Parameter estimation of the homodyned k distribution based on an artificial
  neural network for ultrasound tissue characterization,'' \emph{Ultrasonics},
  vol. 111, p. 106308, 2021.

\bibitem{tehrani2022homodyned}
A.~K. Tehrani, I.~M. Rosado-Mendez, and H.~Rivaz, ``Homodyned k-distribution:
  parameter estimation and uncertainty quantification using bayesian neural
  networks,'' \emph{2023 IEEE 20th International Symposium on Biomedical
  Imaging (ISBI), https://arxiv.org/abs/2211.00175}, 2023.

\bibitem{vincent2008extracting}
P.~Vincent, H.~Larochelle, Y.~Bengio, and P.-A. Manzagol, ``Extracting and
  composing robust features with denoising autoencoders,'' in \emph{Proceedings
  of the 25th international conference on Machine learning}, 2008, pp.
  1096--1103.

\bibitem{ladjal2019pca}
S.~Ladjal, A.~Newson, and C.-H. Pham, ``A pca-like autoencoder,'' \emph{arXiv
  preprint arXiv:1904.01277}, 2019.

\bibitem{wu2023parallelized}
X.~Wu, K.~Lv, S.~Wu, D.-I. Tai, P.-H. Tsui, and Z.~Zhou, ``Parallelized
  ultrasound homodyned-k imaging based on a generalized artificial neural
  network estimator,'' \emph{Ultrasonics}, p. 106987, 2023.

\bibitem{nam2012comparison}
K.~Nam, I.~M. Rosado-Mendez, L.~A. Wirtzfeld, G.~Ghoshal, A.~D. Pawlicki, E.~L.
  Madsen, R.~J. Lavarello, M.~L. Oelze, J.~A. Zagzebski, W.~D. O’Brien~Jr
  \emph{et~al.}, ``Comparison of ultrasound attenuation and backscatter
  estimates in layered tissue-mimicking phantoms among three clinical
  scanners,'' \emph{Ultrasonic imaging}, vol.~34, no.~4, pp. 209--221, 2012.

\bibitem{insana2022acoustic}
M.~F. Insana and D.~G. Brown, ``Acoustic scattering theory applied to soft
  biological tissues,'' in \emph{Ultrasonic scattering in biological
  tissues}.\hskip 1em plus 0.5em minus 0.4em\relax CRC Press, 1993, pp.
  75--124.

\bibitem{bhatt2021multiparametric}
M.~Bhatt, L.~Yazdani, F.~Destrempes, L.~Allard, B.~N. Nguyen, A.~Tang, and
  G.~Cloutier, ``Multiparametric in vivo ultrasound shear wave
  viscoelastography on farm-raised fatty duck livers: human radiology imaging
  applied to food sciences,'' \emph{Poultry science}, vol. 100, no.~4, p.
  100968, 2021.

\bibitem{Tehrani2023}
\BIBentryALTinterwordspacing
A.~K.~Z. Tehrani, I.~M. Rosado-Mendez, H.~Whitson, and H.~Rivaz, ``{A deep
  learning approach for patchless estimation of ultrasound quantitative
  parametric image with uncertainty measurement},'' in \emph{Medical Imaging
  2023: Ultrasonic Imaging and Tomography}, C.~Boehm and N.~Bottenus, Eds.,
  vol. 12470, International Society for Optics and Photonics.\hskip 1em plus
  0.5em minus 0.4em\relax SPIE, 2023, p. 1247010. [Online]. Available:
  \url{https://doi.org/10.1117/12.2651583}
\BIBentrySTDinterwordspacing

\bibitem{roy2018assessment}
M.-H. Roy-Cardinal, F.~Destrempes, G.~Soulez, and G.~Cloutier, ``Assessment of
  carotid artery plaque components with machine learning classification using
  homodyned-k parametric maps and elastograms,'' \emph{IEEE transactions on
  ultrasonics, ferroelectrics, and frequency control}, vol.~66, no.~3, pp.
  493--504, 2018.

\bibitem{mcdonald2009handbook}
J.~H. McDonald, \emph{Handbook of biological statistics}.\hskip 1em plus 0.5em
  minus 0.4em\relax sparky house publishing Baltimore, MD, 2009, vol.~2.

\bibitem{gesnik2020vivo}
M.~Gesnik, M.~Bhatt, M.-H.~R. Cardinal, F.~Destrempes, L.~Allard, B.~N. Nguyen,
  T.~Alquier, J.-F. Giroux, A.~Tang, and G.~Cloutier, ``In vivo ultrafast
  quantitative ultrasound and shear wave elastography imaging on farm-raised
  duck livers during force feeding,'' \emph{Ultrasound in medicine \& biology},
  vol.~46, no.~7, pp. 1715--1726, 2020.

\bibitem{rivaz20079c}
H.~Rivaz, R.~Zellars, G.~Hager, G.~Fichtinger, and E.~Boctor, ``9c-1 beam
  steering approach for speckle characterization and out-of-plane motion
  estimation in real tissue,'' in \emph{2007 IEEE Ultrasonics Symposium
  Proceedings}.\hskip 1em plus 0.5em minus 0.4em\relax IEEE, 2007, pp.
  781--784.

\bibitem{destrempes2021statistical}
F.~Destrempes and G.~Cloutier, ``Statistical modeling of ultrasound signals
  related to the packing factor of wave scattering phenomena for structural
  characterization,'' \emph{The Journal of the Acoustical Society of America},
  vol. 150, no.~5, pp. 3544--3556, 2021.

\bibitem{cloutier2021quantitative}
G.~Cloutier, F.~Destrempes, F.~Yu, and A.~Tang, ``Quantitative ultrasound
  imaging of soft biological tissues: a primer for radiologists and medical
  physicists,'' \emph{Insights into Imaging}, vol.~12, pp. 1--20, 2021.

\end{thebibliography}

	\end{document}


\title{Supplementary Material for \\Homodyned K-Distribution Parameter Estimation in Quantitative Ultrasound: Autoencoder and Bayesian Neural Network Approaches}
	\author{Ali K. Z. Tehrani, Guy Cloutier, An Tang, Ivan M. Rosado-Mendez$^*$, and Hassan Rivaz$^*$ 
		\thanks{A. K. Z. Tehrani and H. Rivaz are with the Department
			of Electrical and Computer Engineering, Concordia University, QC,  Canada.
			An Tang is with the Department of Radiology, Radio-oncology and Nuclear Medicine, University of Montreal, QC, Canada.	
			Guy Cloutier is with the Department of Radiology, Radio-oncology and Nuclear Medicine, and Institute of Biomedical Engineering, University of Montreal, QC, Canada.	
			Ivan M. Rosado-Mendez is with the Department of Medical Physics and Radiology, University of Wisconsin, United States.
			e-mail: A\_Kafaei@encs.concordia.ca, rosadomendez@wisc.edu, an.tang@umontreal.ca, guy.cloutier@umontreal.ca, and  
			hrivaz@ece.concordia.ca. $^*$ represents joint senior authorship with equal contribution.}%
		\thanks{}}

\maketitle
\section{Verifying the implementation of theoretical feature calculation}
\textcolor{black}{In order to verify the correctness of the implementation of theoretical feature values, we generated samples of HK-distribution having a large sample size to ensure the effect of lack of samples is minimized. In the next step, we calculated sample estimate of statistical features from the generated samples, and compared them with the theoretical value. We found a very low value of error which shows that the implementation is correct.}
\section{Simulation Results}
The median, and $25-75\%$ percentile range of simulation results are given in Table \ref{tab:table_sim}.
\section{More experimental phantom results}
\begin{figure}[t]	
	\centering
	\includegraphics[width=0.99\textwidth]{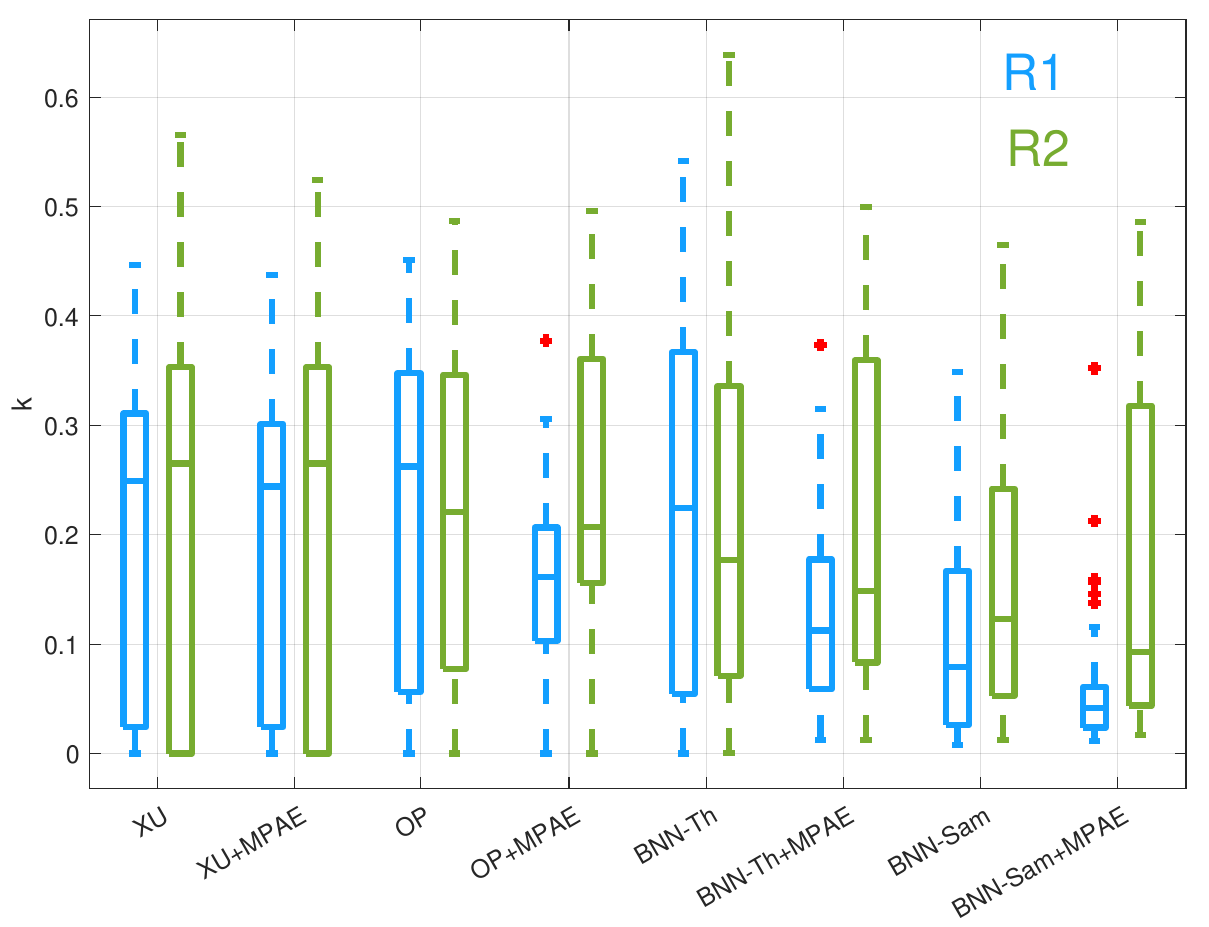}
	\caption{ \textcolor{black}{The boxplot of $k$ of the layered phantom using the evaluated methods. Low coherency is expected for both R1 and R2 regions.} }
	\label{fig:exp_boxplot}	
\end{figure}

The boxplot of the parameter $k$ of R1 and R2 regions of the layered phantom is illustrated in Fig. \ref{fig:exp_boxplot}. Low coherency is expected for both regions.  

\section{MPAE feature reconstruction performance}
\begin{figure}[t]	
	\centering
	\includegraphics[width=0.99\textwidth]{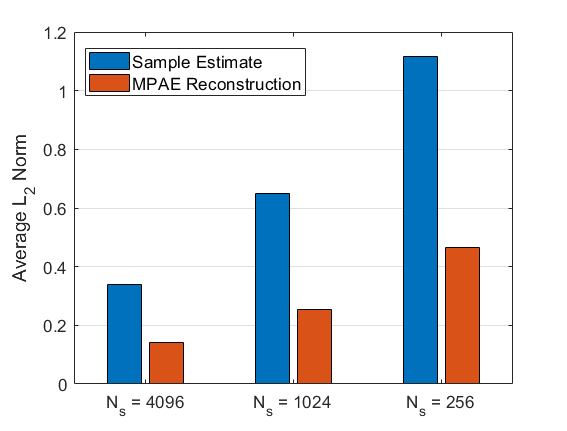}
	\caption{ \textcolor{black}{The average $L_2$ norm of difference between theoretical statistical features and their estimates obtained by sample estimate (blue) and reconstructed features using MPAE (red) for $\rho=0.2$.} }
	\label{fig:feature1}	
\end{figure}

\textcolor{black}{In the manuscript, the estimation performance of MPAE was fully investigated. Here, we evaluated the performance of MPAE in the reconstruction of statistical features. Figures \ref{fig:feature1} illustrates the L2-norm values representing the disparities between the theoretical values of statistical features and the reconstructed features by MPAE (depicted in red), in addition to the sample estimates of features (depicted in blue).}
\section{Simulation results for high correlation}
\begin{figure}[t]	
	\centering
	\includegraphics[width=0.99\textwidth]{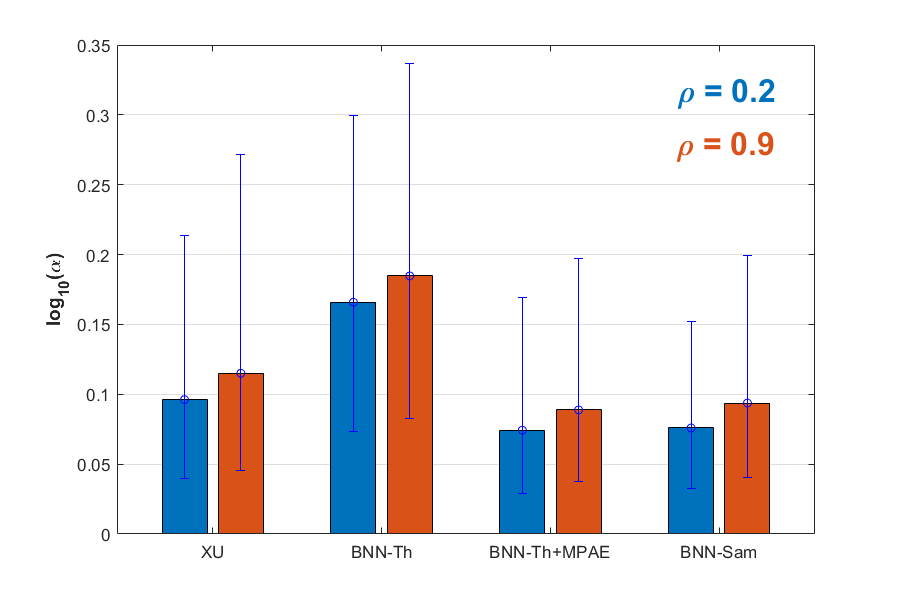}
	\caption{ \textcolor{black}{The median (bar height) and interquartile range (whiskers) of MAE of $log_{10}(\alpha)$ for $\rho=$ 0.2 (blue) and 0.9 (red). The MPAE and BNN-Sam are trained on data with $\rho=$ 0.2.} }
	\label{fig:rho09_1}	
\end{figure}

\begin{figure}[t]	
	\centering
	\includegraphics[width=0.99\textwidth]{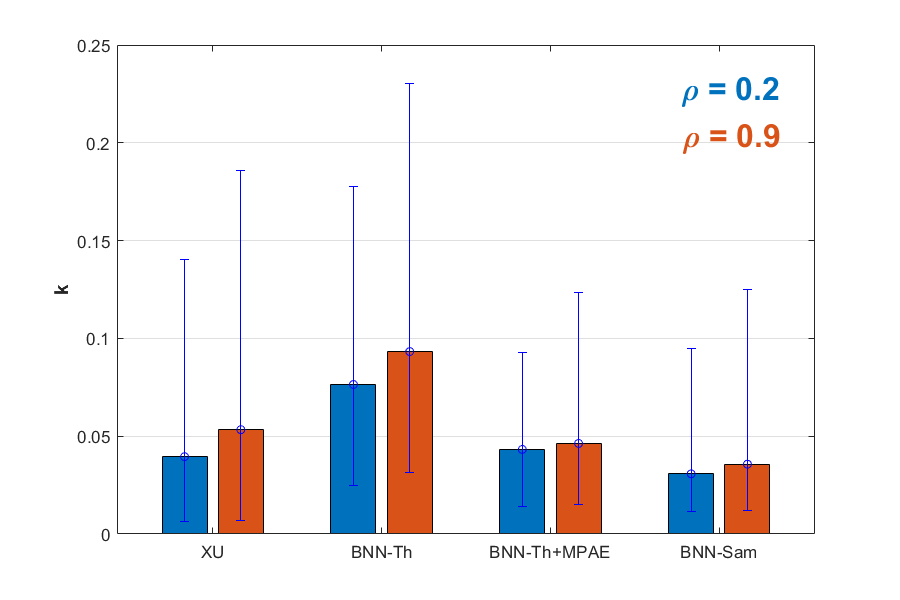}
	\caption{ \textcolor{black}{The median (bar height) and interquartile range (whiskers) of MAE of $k$ for $\rho=$ 0.2 (blue) and 0.9 (red). The MPAE and BNN-Sam are trained on data with $\rho=$ 0.2.} }
	\label{fig:rho09_2}	
\end{figure}

\textcolor{black}{The simulation results are given for a high correlation of $\rho=0.9$ (the MPAE was trained on a low correlation value of $\rho=0.2$) for $log_{10}(\alpha)$ and $k$ in Figs \ref{fig:rho09_1} and \ref{fig:rho09_2}, respectively. It can be observed that the high correlation adversely affects the performance of the estimators.}

\section{Statistical test of experimental phantom data}
\textcolor{black}{The \textit{p}-values of two-sample \textit{F}-test of $log_{10}(\alpha)$ and the ratio of $\alpha$ are reported in Table \ref{tab:phantom_sig} and \ref{tab:ratio_sig}, respectively. In contrast to the statistical test of errors of simulation data,  \textit{F}-test was employed to investigate the similarity of the histograms of estimated values obtained by the evaluated methods, rather than comparing the median values.} 

\section{\textit{in vivo} pre-processing step}
\begin{figure}[t]	
	\centering
	\includegraphics[width=0.7\textwidth]{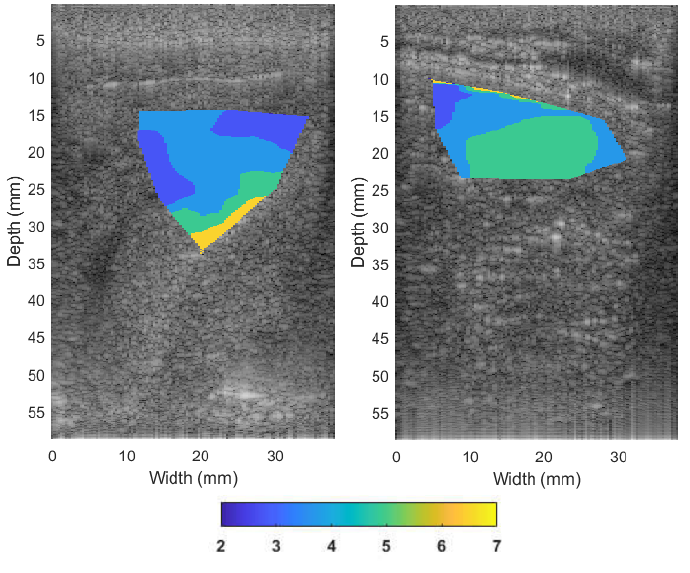}
	\caption{ \textcolor{black}{The segmented regions of duck B (before force feeding: left, and after: right) obtained from the pre-processing method used for \textit{in vivo} data. Statistical features in parametric image are computed from samples belonging to the same class as the center sample.} }
	\label{fig:duck_mask}	
\end{figure}

\textcolor{black}{The pre-processing step is illustrated for duck B in Fig. \ref{fig:duck_mask}. This step ensures that for each patch, only samples belonging to the same class are utilized for statistical feature calculation. A similar strategy was adopted in [37] where an unsupervised method was used to define image pixels into a maximum of three labels (or patches) prior to HK parameter estimation. The number of clusters is a hyper-parameter, and we selected it to be 4.} 

\section{Analysis of $log_{10}(\alpha)$ of the \textit{in vivo} data}
\begin{figure}[t]	
	\centering
	\includegraphics[width=0.85\textwidth]{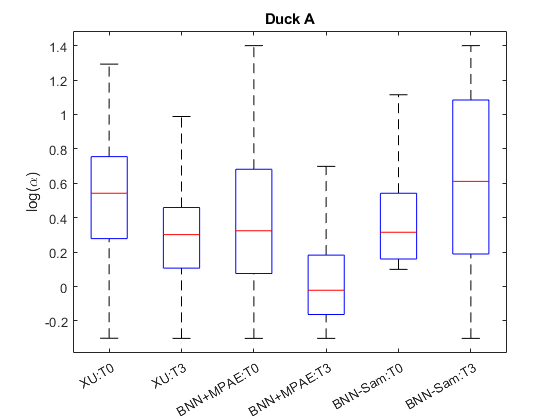}
	\caption{ The box plot of values of parametric images of $log_{10}(\alpha)$ of Duck A before force feeding (T0), and after (T3). All estimated $log_{10}(\alpha)$ values within the liver for one frame were employed to calculate this plot.}
	\label{fig:ducka2}	
\end{figure}

\begin{figure}[t]	
	\centering
	\includegraphics[width=0.85\textwidth]{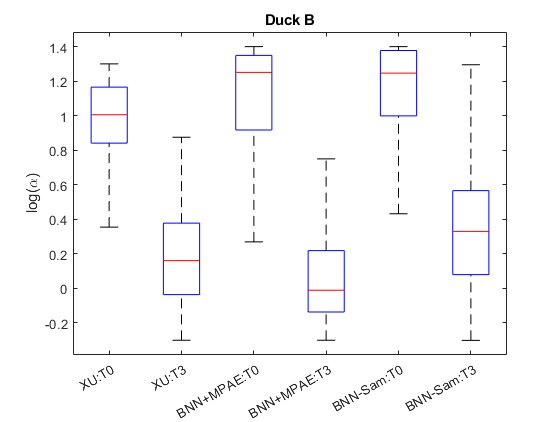}
	\caption{ The box plot of values of parametric image of $log_{10}(\alpha)$ of Duck B before force feeding (T0), and after (T3). All estimated $log_{10}(\alpha)$ values within the liver for one frame were employed to calculate this plot.}
	\label{fig:duckb2}	
\end{figure}

\begin{figure}[t]	
	\centering
	\includegraphics[width=0.85\textwidth]{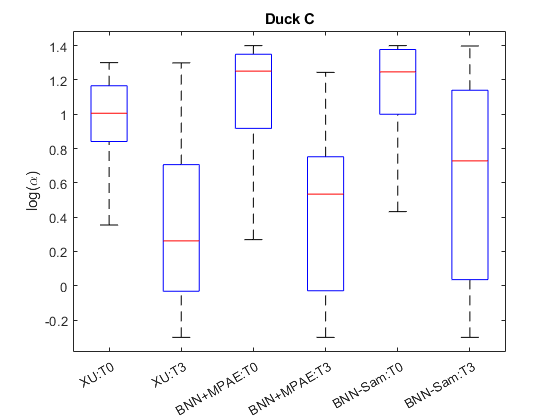}
	\caption{ The box plot of values of parametric image of $log_{10}(\alpha)$ of Duck C before force feeding (T0), and after (T3). All estimated $log_{10}(\alpha)$ values within the liver for one frame were employed to calculate this plot.}
	\label{fig:duckc2}	
\end{figure}

\textcolor{black}{the box plot of $log_{10}(\alpha)$ values of the three ducks are illustrated in Fig. \ref{fig:ducka2}, \ref{fig:duckb2}  and \ref{fig:duckc2}. Each box summarizes the statistics from the values extracted from the region of interest defined in one frame of the liver at each time point.}

\section{More parametric images}
\begin{figure}[t]	
	\centering
	\includegraphics[width=0.99\textwidth]{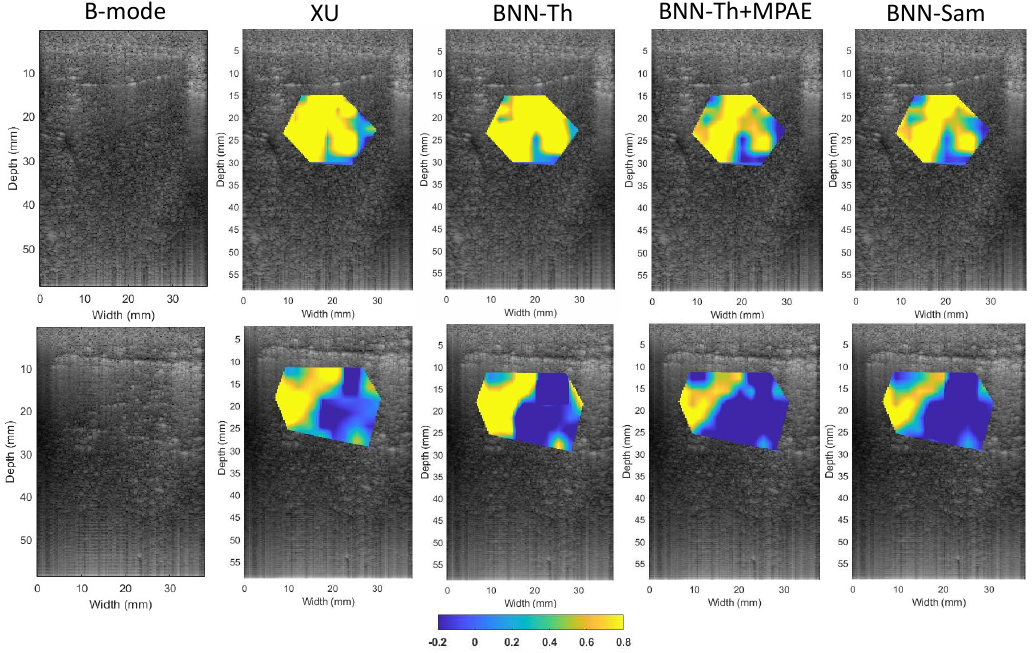}
	\caption{ The B-mode image (first column) and the parametric image of $log_{10}(\alpha)$ of duck C before force feeding (top row), and after (bottom row).}
	\label{fig:DA_T0_param}	
\end{figure}

\begin{figure}[t]	
	\centering
	\includegraphics[width=0.98\textwidth]{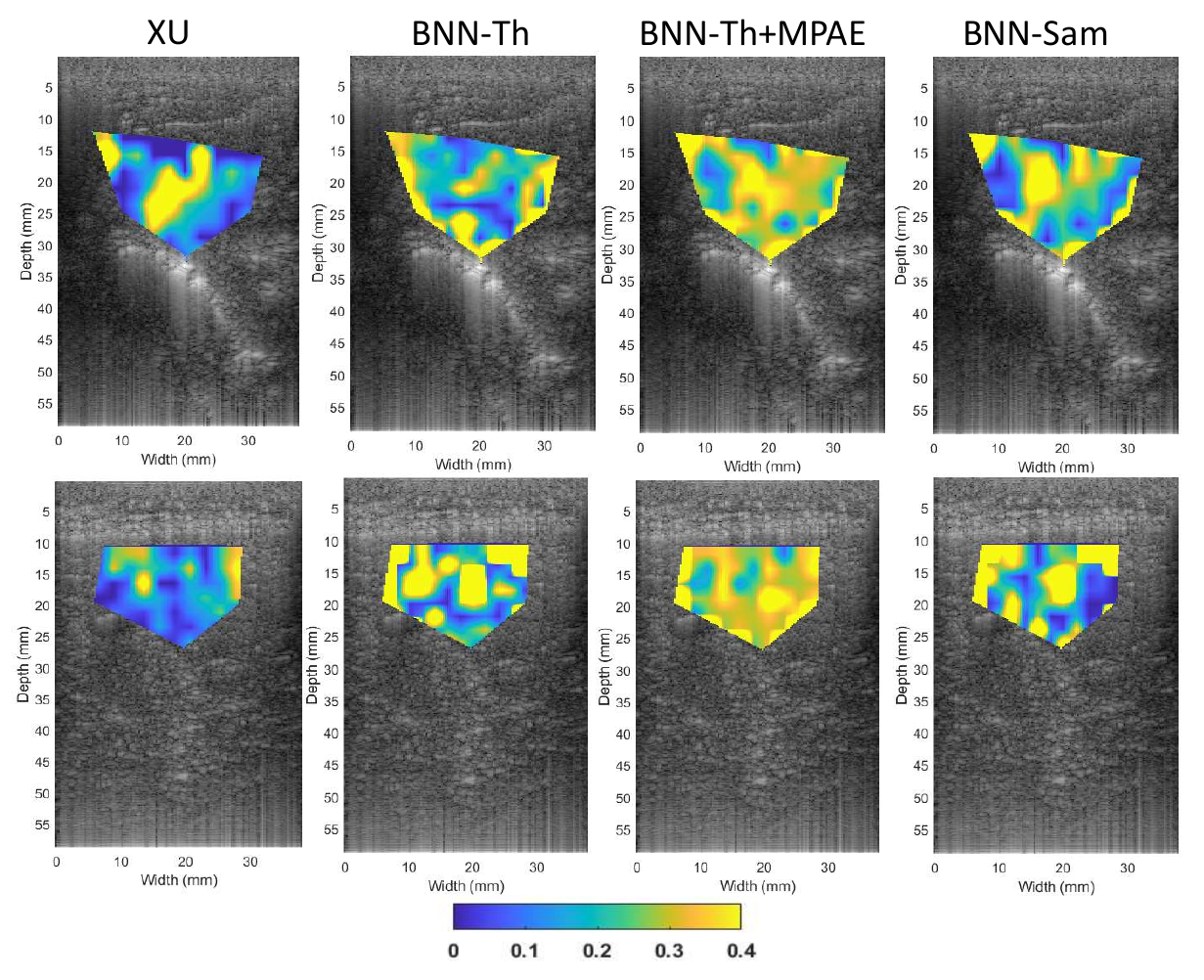}
	\caption{ The parametric image of $k$ of duck A before force feeding (top row), and after (bottom row). }
	\label{fig:DA_k}	
\end{figure}

\begin{figure}[t]	
	\centering
	\includegraphics[width=0.98\textwidth]{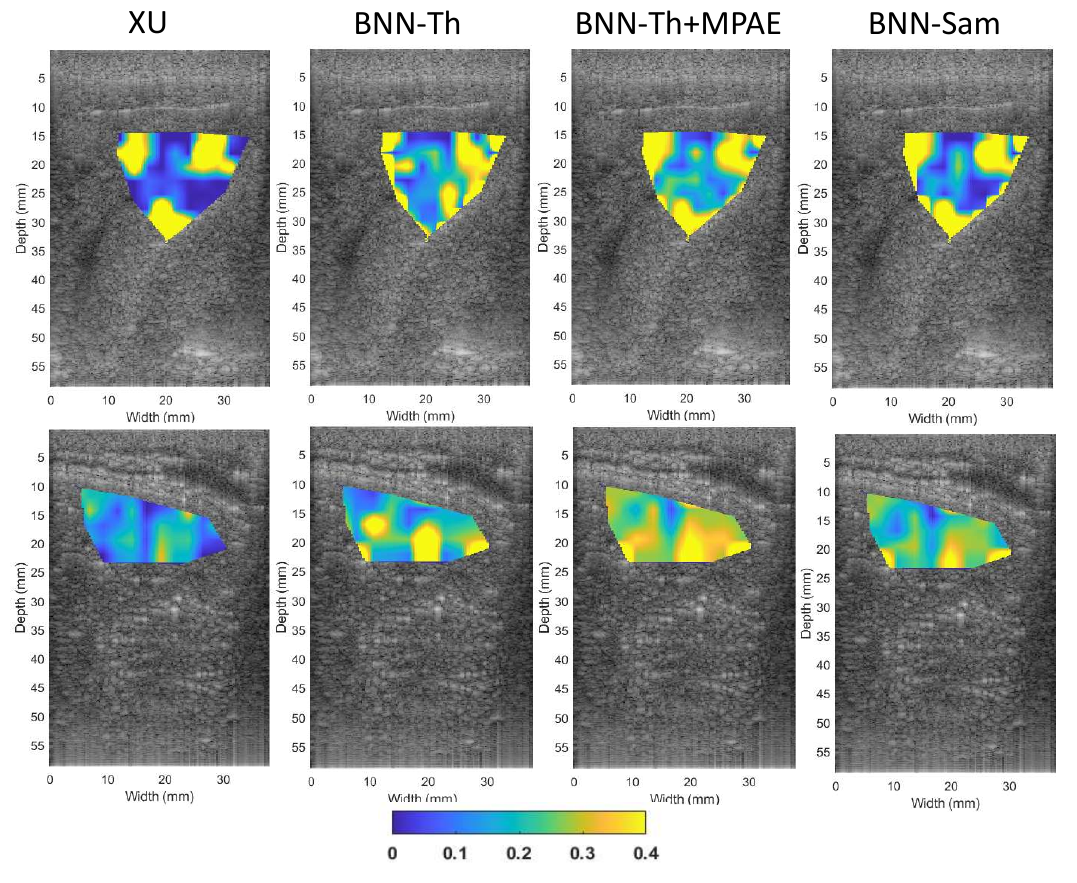}
	\caption{ The parametric image of $k$ of duck B before force feeding (top row), and after (bottom row). }
	\label{fig:DB_k}	
\end{figure}

The parametric image of $log_{10}(\alpha)$ of duck C is shown in Fig. \ref{fig:DA_T0_param}. The parametric images of $k$ of duck A and B are shown in Fig. \ref{fig:DA_k} and Fig. \ref{fig:DB_k}.

\begin{table*}[]
	\caption{Simulation results are reported by median \small[$25\%$,$75\%$]. The lowest median error for each sample size is marked in bold.}
	\label{tab:table_sim}
	\begin{tabular}{@{}cccccc@{}}
		\toprule
		Method & Sample Size & MAE of $log_{10}(\alpha)$ & RRMSE of $log_{10}(\alpha)$ & MAE of $k$ & RRMSE of $k$ \\ \midrule
		XU & 4096 & 0.050 {[}0.021,0.130{]} & 0.128 {[}0.054,0.247{]} & 0.024 {[}0.005,0.082{]} & \textbf{0.128} {[}0.032,0.548{]} \\
		\rowcolor[HTML]{EFEFEF} 
		XU & 1024 & 0.096 {[}0.040,0.213{]} & 0.229 {[}0.101,0.392{]} & 0.040 {[}0.006,0.140{]} & \textbf{0.228} {[}0.059,0.752{]} \\
		XU & 256 & 0.166 {[}0.072,0.354{]} & 0.378 {[}0.175,0.629{]} & 0.076 {[}0.013,0.225{]} & 0.398 {[}0.093,0.962{]} \\ \midrule
		\rowcolor[HTML]{EFEFEF} 
		XU+MPAE & 4096 & 0.048 {[}0.020,0.132{]} & 0.123 {[}0.052,0.239{]} & 0.029 {[}0.005,0.085{]} & 0.140 {[}0.040,0.528{]} \\
		XU+MPAE & 1024 & 0.088 {[}0.036,0.207{]} & 0.206 {[}0.093,0.363{]} & 0.040 {[}0.006,0.140{]} & 0.232 {[}0.062,0.752{]} \\
		\rowcolor[HTML]{EFEFEF} 
		XU+MPAE & 256 & 0.152 {[}0.064,0.325{]} & 0.341 {[}0.161,0.555{]} & 0.065 {[}0.012,0.193{]} & 0.327 {[}0.093,0.880{]} \\ \midrule
		OP & 4096 & 0.070 {[}0.029,0.143{]} & 0.153 {[}0.068,0.277{]} & 0.090 {[}0.028,0.262{]} & 0.257 {[}0.056,0.846{]} \\
		\rowcolor[HTML]{EFEFEF} 
		OP & 1024 & 0.123 {[}0.054,0.234{]} & 0.251 {[}0.122,0.457{]} & 0.122 {[}0.044,0.320{]} & 0.463 {[}0.101,0.888{]} \\
		OP & 256 & 0.198 {[}0.093,0.372{]} & 0.401 {[}0.179,0.741{]} & 0.190 {[}0.067,0.445{]} & 0.655 {[}0.167,0.934{]} \\ \midrule
		\rowcolor[HTML]{EFEFEF} 
		OP+MPAE & 4096 & 0.042 {[}0.017,0.108{]} & 0.100 {[}0.046,0.182{]} & 0.081 {[}0.030,0.222{]} & 0.222 {[}0.058,0.981{]} \\
		OP+MPAE & 1024 & 0.077 {[}0.029,0.172{]} & 0.165 {[}0.080,0.276{]} & 0.121 {[}0.046,0.280{]} & 0.299 {[}0.092,1.530{]} \\
		\rowcolor[HTML]{EFEFEF} 
		OP+MPAE & 256 & 0.198 {[}0.092,0.372{]} & 0.401 {[}0.178,0.741{]} & 0.190 {[}0.068,0.442{]} & 0.643 {[}0.167,0.934{]} \\ \midrule
		BNN-Th & 4096 & 0.095 {[}0.040,0.180{]} & 0.197 {[}0.092,0.348{]} & 0.040 {[}0.014,0.102{]} & 0.272 {[}0.089,0.818{]} \\
		\rowcolor[HTML]{EFEFEF} 
		BNN-Th & 1024 & 0.166 {[}0.074,0.300{]} & 0.326 {[}0.154,0.602{]} & 0.076 {[}0.025,0.178{]} & 0.477 {[}0.161,1.611{]} \\
		BNN-Th & 256 & 0.283 {[}0.134,0.480{]} & 0.515 {[}0.250,1.071{]} & 0.147 {[}0.050,0.305{]} & 0.817 {[}0.302,3.201{]} \\ \midrule
		\rowcolor[HTML]{EFEFEF} 
		BNN-Th+MPAE & 4096 & 0.044 {[}0.017,0.105{]} & 0.098 {[}0.046,0.180{]} & 0.026 {[}0.008,0.060{]} & 0.173 {[}0.052,0.534{]} \\
		BNN-Th+MPAE & 1024 & 0.074 {[}0.029,0.170{]} & 0.163 {[}0.078,0.270{]} & 0.043 {[}0.014,0.093{]} & 0.274 {[}0.090,0.743{]} \\
		\rowcolor[HTML]{EFEFEF} 
		BNN-Th+MPAE & 256 & 0.133 {[}0.057,0.254{]} & 0.259 {[}0.134,0.448{]} & 0.081 {[}0.024,0.169{]} & 0.444 {[}0.167,1.553{]} \\ \midrule
		BNN-Sam & 4096 & 0.043 {[}0.017,0.101{]} & 0.095 {[}0.042,0.178{]} & \textbf{0.020} {[}0.007,0.060{]} & 0.171 {[}0.058,0.389{]} \\
		\rowcolor[HTML]{EFEFEF} 
		BNN-Sam & 1024 & 0.076 {[}0.033,0.152{]} & 0.153 {[}0.072,0.302{]} & 0.031 {[}0.012,0.095{]} & 0.278 {[}0.119,0.531{]} \\
		BNN-Sam & 256 & 0.128 {[}0.058,0.238{]} & 0.239 {[}0.119,0.477{]} & \textbf{0.045} {[}0.015,0.129{]} & 0.381 {[}0.165,0.644{]} \\ \midrule
		\rowcolor[HTML]{EFEFEF} 
		BNN-Sam+MPAE & 4096 & \textbf{0.041} {[}0.017,0.097{]} & \textbf{0.090} {[}0.041,0.165{]} & \textbf{0.020} {[}0.007,0.053{]} & 0.168 {[}0.054,0.402{]} \\
		BNN-Sam+MPAE & 1024 & \textbf{0.070} {[}0.028,0.149{]} & \textbf{0.143} {[}0.067,0.268{]} & \textbf{0.026} {[}0.010,0.081{]} & 0.244 {[}0.086,0.484{]} \\
		\rowcolor[HTML]{EFEFEF} 
		BNN-Sam+MPAE & 256 & \textbf{0.123} {[}0.054,0.234{]} & \textbf{0.227} {[}0.118,0.452{]} & 0.047 {[}0.015,0.130{]} & \textbf{0.377} {[}0.154,0.657{]} \\ \bottomrule
	\end{tabular}
\end{table*}

\newpage 

\begin{table}[]
	\caption{\textit{p}-values of  two-sample \textit{F}-test for $log_{10}(\alpha)$ of the layered phantom. Only pairs with \textit{p}-values$>0.001$ are reported.}
	\label{tab:phantom_sig}
	\begin{tabular}{@{}ccc@{}}
		\toprule
		Pairs & Region & \textit{p}-value \\ \midrule
		\rowcolor[HTML]{EFEFEF} 
		XU, XU+MPAE & R1 & 0.723 \\
		XU,OP & R1 & 0.307 \\
		\rowcolor[HTML]{EFEFEF} 
		XU, BNN-Th & R1 & 0.029 \\
		XU, BNN-Sam & R1 & 0.144 \\
		\rowcolor[HTML]{EFEFEF} 
		XU, BNN-Sam+MPAE & R1 & 0.012 \\
		XU+MPAE, OP & R1 & 0.17 \\
		\rowcolor[HTML]{EFEFEF} 
		XU+MPAE, BNN-Th & R1 & 0.011 \\
		XU+MPAE, BNN+MPAE & R1 & 0.002 \\
		\rowcolor[HTML]{EFEFEF} 
		XU+MPAE, BNN-Sam & R1 & 0.267 \\
		XU+MPAE, BNN-Sam+MPAE & R1 & 0.03 \\ \midrule
		\rowcolor[HTML]{EFEFEF} 
		OP, BNN-Th & R1 & 0.239 \\
		OP, BNN-Sam & R1 & 0.014 \\
		\rowcolor[HTML]{EFEFEF} 
		OP+MPAE, BNN-Th+MPAE & R1 & 0.849 \\
		OP+MPAE, BNN-Sam & R1 & 0.032 \\
		\rowcolor[HTML]{EFEFEF} 
		OP+MPAE, BNN-Sam+MPAE & R1 & 0.278 \\ \midrule
		BNN+MPAE, BNN-Sam & R1 & 0.050 \\
		\rowcolor[HTML]{EFEFEF} 
		BNN+MPAE, BNN-Sam+MPAE & R1 & 0.370 \\
		BNN-Sam, BNN-Sam+MPAE & R1 & 0.283 \\ \midrule
		\rowcolor[HTML]{EFEFEF} 
		XU, XU+MPAE & R2 & 0.400 \\
		XU, OP & R2 & 0.203 \\
		\rowcolor[HTML]{EFEFEF} 
		XU, BNN-Th & R2 & 0.374 \\
		XU+MPAE, OP & R2 & 0.036 \\
		\rowcolor[HTML]{EFEFEF} 
		XU+MPAE, BNN-Th & R2 & 0.086 \\ \midrule
		OP, BNN-Th & R2 & 0.700 \\
		\rowcolor[HTML]{EFEFEF} 
		OP+MPAE, BNN-Th+MPAE & R2 & 0.948 \\
		OP+MPAE, BNN-Sam & R2 & 0.233 \\
		\rowcolor[HTML]{EFEFEF} 
		OP+MPAE, BNN-Sam+MPAE & R2 & 0.200 \\ \midrule
		BNN-Th+MPAE, BNN-Sam & R2 & 0.209 \\
		\rowcolor[HTML]{EFEFEF} 
		BNN-Th+MPAE, BNN-Sam+MPAE & R2 & 0.179 \\ \midrule
		BNN-Sam, BNN-Sam+MPAE & R2 & 0.931 \\ \bottomrule
	\end{tabular}
\end{table}

\begin{table}[]
	\caption{\textit{p-}values of  two-sample \textit{F}-test  for the ratio of $\alpha$ of the layered phantom. Only pairs with \textit{p-}values$>0.001$ are reported.}
	\label{tab:ratio_sig}
	\begin{tabular}{@{}cc@{}}
		\toprule
		Pairs & \textit{p-}values \\ \midrule
		XU, XU+MPAE & 0.120 \\
		\rowcolor[HTML]{EFEFEF} 
		XU, OP & 0.860 \\
		XU, BNN-Th & 0.522 \\ \midrule
		\rowcolor[HTML]{EFEFEF} 
		XU+MPAE, BNN-Th & 0.029 \\ \midrule
		OP, BNN-Th & 0.415 \\ \midrule
		\rowcolor[HTML]{EFEFEF} 
		OP+MPAE, BNN-Th+MPAE & 0.828 \\
		OP+MPAE, BNN-Th+MPAE & 0.015 \\
		\rowcolor[HTML]{EFEFEF} 
		OP+MPAE, BNN-Sam+MPAE & 0.026 \\ \midrule
		BNN-Th+MPAE, BNN-Sam+MPAE & 0.921 \\ \midrule
		\rowcolor[HTML]{EFEFEF} 
		BNN-Sam, BNN-Sam+MPAE & 0.033 \\ \bottomrule
	\end{tabular}
\end{table}


